\documentclass[twoside,twocolumn,9pt]{article}
\usepackage{extsizes}
\usepackage[super,sort&compress,comma]{natbib} 
\usepackage[version=3]{mhchem}
\usepackage[left=1.5cm, right=1.5cm, top=1.785cm, bottom=2.0cm]{geometry}
\usepackage{balance}
\usepackage{times,mathptmx}
\usepackage{sectsty}
\usepackage{graphicx} 
\usepackage{lastpage}
\usepackage[format=plain,justification=justified,singlelinecheck=false,font={stretch=1.125,small,sf},labelfont=bf,labelsep=space]{caption}
\usepackage{float}
\usepackage{fnpos}
\usepackage[english]{babel}
\usepackage{array}
\usepackage{droidsans}
\usepackage{charter}
\usepackage[T1]{fontenc}
\usepackage[usenames,dvipsnames]{xcolor}
\usepackage{setspace}
\usepackage[compact]{titlesec}
\usepackage{hyperref}

\usepackage{braket}
\usepackage{amsmath} 
\usepackage{amssymb} 
\usepackage{bm} 

\begin{document}


\makeFNbottom
\makeatletter
\renewcommand\LARGE{\@setfontsize\LARGE{15pt}{17}}
\renewcommand\Large{\@setfontsize\Large{12pt}{14}}
\renewcommand\large{\@setfontsize\large{10pt}{12}}
\renewcommand\footnotesize{\@setfontsize\footnotesize{7pt}{10}}
\makeatother

\renewcommand{\thefootnote}{\fnsymbol{footnote}}
\renewcommand\footnoterule{\vspace*{1pt}%
\color{black}\hrule width 3.5in height 0.4pt \color{black}\vspace*{5pt}} 
\setcounter{secnumdepth}{5}

\makeatletter 
\renewcommand\@biblabel[1]{#1}            
\renewcommand\@makefntext[1]%
{\noindent\makebox[0pt][r]{\@thefnmark\,}#1}
\makeatother 
\renewcommand{\figurename}{\small{Fig.}~}
\sectionfont{\sffamily\Large}
\subsectionfont{\normalsize}
\subsubsectionfont{\bf}
\setstretch{1.125} 
\setlength{\skip\footins}{0.8cm}
\setlength{\footnotesep}{0.25cm}
\setlength{\jot}{10pt}
\titlespacing*{\section}{0pt}{4pt}{4pt}
\titlespacing*{\subsection}{0pt}{15pt}{1pt}

\setlength{\arrayrulewidth}{1pt}
\setlength{\columnsep}{6.5mm}
\setlength\bibsep{1pt}

\makeatletter
\newcommand{\topfigrule}{\vspace*{-1pt}}
\newcommand{\botfigrule}{\vspace*{-2pt}}
\newcommand{\dblfigrule}{\vspace*{-1pt}}
\makeatother

\newcommand{\onlinecite}[1]{\hspace{-1 ex} \nocite{#1}\citenum{#1}} 

\twocolumn[
  \begin{@twocolumnfalse}
\vspace{3cm}

\begin{tabular}{m{2.5cm} p{13.5cm} }

&\noindent\LARGE{\textbf{Addressing the exciton fine structure in colloidal nanocrystals: the case of CdSe nanoplatelets}} \\
\vspace{0.3cm} & \vspace{0.3cm} \\

 &  \noindent\large{Elena V. Shornikova,$^{\ast}$\textit{$^{a,b}$} Louis Biadala,$^{\ast}$\textit{$^{a,c}$} Dmitri R. Yakovlev,$^{\ast}$\textit{$^{a,d}$}\textit{$^{\ddag}$} Victor F. Sapega,\textit{$^{d}$} Yuri G. Kusrayev,\textit{$^{d}$} Anatolie A. Mitioglu,\textit{$^{e}$} Mariana V. Ballottin,\textit{$^{e}$} Peter C. M. Christianen,\textit{$^{e}$} Vasilii V. Belykh,\textit{$^{a,f}$} Mikhail V. Kochiev,\textit{$^{f}$} Nikolai N. Sibeldin,\textit{$^{f}$} Aleksandr A. Golovatenko,\textit{$^{d}$} Anna V. Rodina,\textit{$^{d}$} Nikolay A. Gippius,\textit{$^{g}$}  Alexis Kuntzmann,\textit{$^{h}$} Ye Jiang,\textit{$^{h}$} Michel Nasilowski,\textit{$^{h}$} Benoit Dubertret,\textit{$^{h}$} and Manfred Bayer\textit{$^{a,d}$}} \\

\vspace{1 cm}\\
& \noindent\normalsize{
We study the band-edge exciton fine structure and in particular its bright-dark splitting in colloidal semiconductor nanocrystals by four different optical methods based on fluorescence line narrowing and time-resolved measurements at various temperatures down to 2 K. We demonstrate that all these methods provide consistent splitting values and discuss their advances and limitations. Colloidal CdSe nanoplatelets with thicknesses of 3, 4 and 5 monolayers are chosen for experimental demonstrations. The bright-dark splitting of excitons varies from 3.2 to 6.0~meV and is inversely proportional to the nanoplatelet thickness. Good agreement between experimental and theoretically calculated size dependence of the bright-dark exciton slitting is achieved. The recombination rates of the bright and dark excitons and the bright to dark relaxation rate are measured by time-resolved techniques.} \\

\end{tabular}

 \end{@twocolumnfalse} \vspace{0.6cm}

  ]

\renewcommand*\rmdefault{bch}\normalfont\upshape
\rmfamily
\section*{}
\vspace{-1cm}


\footnotetext{\textit{$^{a}$~Experimentelle Physik 2, Technische Universit{\"a}t Dortmund, 44221 Dortmund, Germany. Tel: +49 231 755 3531; E-mail: elena.kozhemyakina@tu-dortmund.de, dmitri.yakovlev@tu-dortmund.de}}
\footnotetext{\textit{$^{b}$~Rzhanov Institute of Semiconductor Physics, Siberian Branch of Russian Academy of Sciences, 630090 Novosibirsk, Russia.}}
\footnotetext{\textit{$^{c}$~Institut d'Electronique, de Micro{\'e}lectronique et de Nanotechnologie, CNRS, 59652 Villeneuve-d'Ascq, France. Tel:  +33 3 20 19 79 32; E-mail: louis.biadala@isen.iemn.univ-lille1.fr}}
\footnotetext{\textit{$^{d}$~Ioffe  Institute, Russian Academy of Sciences, 194021 St. Petersburg, Russia. }}
\footnotetext{\textit{$^{e}$~High Field Magnet Laboratory (HFML-EMFL), Radboud University, 6525 ED Nijmegen, The Netherlands.}}
\footnotetext{\textit{$^{f}$~P. N. Lebedev Physical Institute, Russian Academy of Sciences, 119991 Moscow, Russia.}}
\footnotetext{\textit{$^{g}$~Skolkovo Institute of Science and Technology, 143026 Moscow, Russia.}}
\footnotetext{\textit{$^{h}$~Laboratoire de Physique et d'Etude des Mat\'{e}riaux, ESPCI, CNRS, 75231 Paris, France.}}

\section{Introduction}
Colloidal nanostructures are intensively investigated because of their bright luminescence and simplicity of fabrication. Starting from 1993\cite{Murray1993}, the research has been concentrated on nanometer-sized spherical nanocrystals (NCs), also known as quantum dots (QDs).
Recently, two-dimensional nanoplatelets (NPLs) have been synthesized and have attracted great attention due to their remarkable properties. Most importantly, CdSe NPLs with zinc-blende crystal structure have short spontaneous recombination rates,\cite{Ithurria2011nm} narrow ensemble emission spectra due to their atomically controlled thickness,\cite{Ithurria2008} and dipole emission oriented within the plane.\cite{Gao2017} Among other important properties the very efficient fluorescence resonance energy transfer,\cite{Rowland2015} the ultralow stimulated emission threshold,\cite{Grim2014,Diroll2017} the enhanced conductivity due to in-plane transport,
\cite{Zhang2005,Schliehe2010, Dogan2015} and the highly efficient charge carrier multiplication\cite{Aerts2014} can be highlighted. Widely varying structures have been synthesized: CdSe wurtzite nanoribbons or quantum belts\cite{Joo2006, Son2009, Liu2010}, NPLs of PbS,\cite{Schliehe2010,Dogan2015} PbSe,\cite{Koh2017} Cu$_{2-x}$S\cite{Sigman2003, Zhang2005}, GeS and GeSe,\cite{VaughnII2010} CdS,\cite{Ithurria2011nm, Li2012} ZnS\cite{Bouet2014}, CdTe\cite{Ithurria2011nm}, and HgTe\cite{Izquierdo2016}, as well as various core-shell structures (for a review see Ref.~\onlinecite{Nasilowski2016}). Among them CdSe-based NPLs play a role of the model system, which optical properties including quantum coherence and exciton dephasing have been intensively studied.\cite{Yeltik2015,Achtstein2015,Cassette2015,Pal2017} 
Compared to bulk semiconductors, in NPLs the exciton binding energy is drastically increased, \textit{e.g.} in CdSe from 10 meV to hundreds of meV. There are three reasons for this:  (i) the large electron effective mass due to nonparabolicity of the conduction band, (ii) the dimensionality reduction, and (iii) the dielectric confinement.\cite{Benchamekh2014} This raises questions about the band-edge exciton fine structure and exciton recombination dynamics in these two-dimensional nanostructures.

Similar to CdSe QDs, the exciton ground state in CdSe NPLs is a two-fold degenerate dark state $\ket{F}$ with angular momentum projections $\pm 2$ on the quantization axis.\cite{Biadala2014nl} The first excited state with angular momentum projection $\pm 1$ is an optically active (bright) $\ket{A}$ state, which is separated from the ground state by a bright-dark energy splitting ($\Delta E_{\rm AF}$) of several meV. The direct observation of the fine structure states in an ensemble of NCs is often hindered by the line broadening resulting from size dispersion. Typically, the linewidth of ensemble photoluminescence (PL) spectra is in the 100~meV range, much larger than $\Delta E_{\rm AF}$ of $1-20$~meV. Two optical methods are commonly used to measure $\Delta E_{\rm AF}$. The first technique is based on fluorescence line narrowing (FLN), which gives direct access to $\Delta E_{\rm AF}$.\cite{GranadosDelAguila2014} The second method relies on the evaluation of $\Delta E_{\rm AF}$ from the temperature dependence of the PL decay\cite{Labeau2003,Biadala2009} (more details are given in Supplementary Section~S1). While these two methods gave a similar result being applied to the same CdSe/CdS core/shell QDs with a 3~nm core diameter,\cite{Brovelli2011} no comparison has been made on the same bare core NCs. It is important to do as a large discrepancy can be found in literature for QDs with diameters less than 3~nm. Nirmal \textit{et al.} measured 19~meV in 2.4~nm diameter bare core CdSe QDs by the FLN technique,\cite{Nirmal1995,Efros1996} while de~Mello~Donega \textit{et al.} reported $\Delta E_{\rm AF} = 1.7$~meV in bare core CdSe QDs with diameter of 1.7~nm from temperature-dependent time-resolved PL,\cite{DeMelloDonega2006} claiming that FLN measurements systematically overestimate $\Delta E_{\rm AF}$ by neglecting any internal relaxation between the exciton states. Recently, it was shown that the Stokes shift in bare core CdSe QDs can be also contributed  by formation of a dangling bond magnetic polaron.\cite{Biadala2017nn,Rodina2015} Moreover, $\Delta E_{\rm AF}$ in QDs is strongly affected by the dot shape and symmetry,\cite{Efros1996,Leung1998} which complicates the comparison of results obtained by different groups. Obviously, more experimental methods are very welcomed to address the measurements of the bright-dark exciton splitting in colloidal nanostructures. Here we suggest and test a few new experimental approaches and examine them together with the commonly used ones on the same samples of CdSe NPLs.

In this paper, we exploit four optical methods to study the bright-dark exciton energy splitting in ensemble measurements of CdSe nanoplatelets with thicknesses ranging from 3 to 5 monolayers: (i) fluorescence line narrowing, (ii) temperature-dependent time-resolved PL, (iii) spectrally-resolved PL decay at cryogenic temperatures, and (iv) temperature dependence of PL spectra. Most importantly, we compare fluorescence line narrowing and temperature-dependent time-resolved PL techniques applied to all samples. The results gained by different methods are in good agreement with each other and confirm the bright-dark exciton splitting of several meV in CdSe NPLs measured earlier by one of the methods.\cite{Biadala2014nl} Comparison of the thickness dependence of the splitting with the results of model calculations allows us to estimate the exchange strength constant and the dielectric constants inside and outside the nanoplatelets. Theoretical calculations within the effective mass approximation with account of the dielectric effect successfully reproduce experimental size dependence of the bright-dark exciton splitting.




\section{Experimental results}

The investigated samples are three batches of CdSe NPLs with thicknesses of $L=3$, 4, and 5 monolayers (MLs) of CdSe and an additional layer of Cd atoms, so that both sides of the NPLs are Cd-terminated. In the following, these samples will be accordingly referred to as 3ML, 4ML, and 5ML. TEM images of the samples are shown in Fig.~\ref{fig:TEM}. Parameters of the studied samples are summarized in Table~\ref{tab:table1}.

\begin{table}[h]
	\small
	\caption{\ Parameters of CdSe nanoplatelets}
	\begin{tabular*}{0.48\textwidth}{@{\extracolsep{\fill}}llll}
		\hline
		Sample & 3ML & 4ML & 5ML \\
		\hline
		Thickness $L$, monolayers & 3 & 4 & 5 \\
		Thickness $L$, nm &  0.9  & 1.2 & 1.5 \\
		Lateral dimensions, nm$^2$ & $6 \times 40$ & $8 \times 16$ & $7 \times 30$ \\
		Emission wavelength ($T=300$~K), nm & 456 & 512 & 551 \\
		Emission photon energy ($T=300$~K), eV & 2.717 & 2.420 & 2.248  \\
		Emission photon energy ($T=4$~K), eV & 2.804 & 2.497 & 2.319  \\
		FWHM of exciton line ($T=300$~K), meV & 50 & 44 & 40  \\
		FWHM of exciton line ($T=4$~K), meV & 19.6 & 16.5 & 15  \\
		Shift between emission lines ($T=4$~K), meV & 30 & 20 & 18  \\
		Light-heavy hole splitting, meV & 159 & 157 & 137 \\
		\hline
	\end{tabular*}
	\label{tab:table1}
\end{table}

\begin{figure*}[h!]
	\centering
	\includegraphics[width=16 cm]{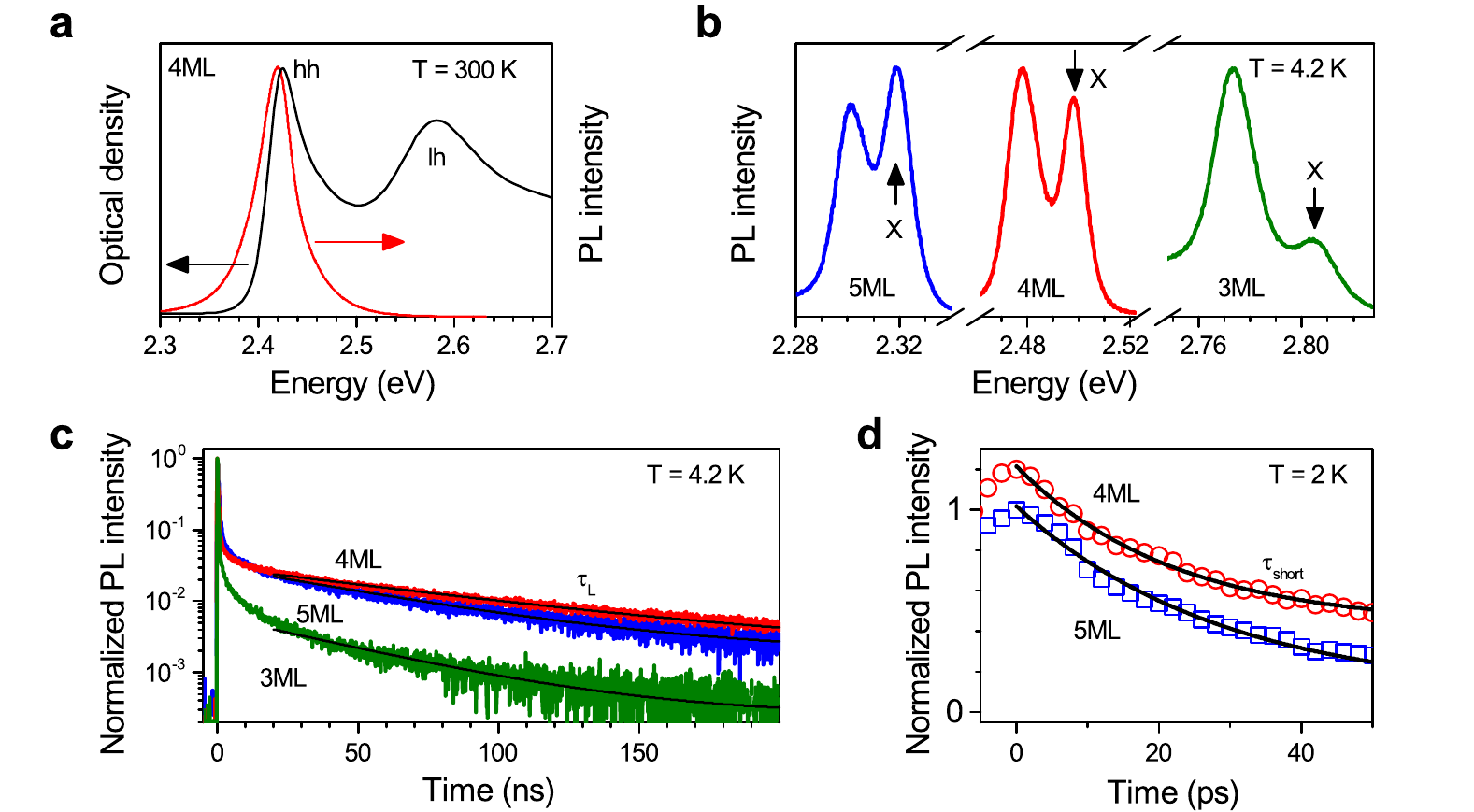}
	\caption{(a) Photoluminescence (red) and absorption (black) spectra of 4ML CdSe NPLs at room temperature. (b) PL spectra of 3ML, 4ML, and 5ML samples at $T=4.2$~K. Exciton peaks are marked by the arrows.  (c) PL decays of 3ML (green), 4ML (red), and 5ML (blue) samples measured at the exciton peaks with an APD. Black lines are single exponential fits with $\tau_{\rm L}$ given in Table~\ref{tab:table_model1}. (d) Normalized PL decays of 4ML (red symbols) and 5ML (blue symbols) samples recorded with a streak camera. Data for 4ML sample are shifted vertically for clarity. Temporal resolution is $\lesssim 5$~ps (Experimental section). Black lines are single exponential fits with $\tau_{\rm short}$ given in Table~\ref{tab:table_model1}.}
	\label{fig:PL}
\end{figure*}

Room temperature PL and absorption spectra of the 4ML NPLs are shown in Fig.~\ref{fig:PL}a. In absorption spectra two peaks at 2.426 eV and 2.583 eV, separated from each other by 157~meV, are related to excitons involving the heavy hole (hh) and light hole (lh), respectively. The heavy-hole exciton has a narrow emission line at room temperature with full width at half maximum (FWHM) of 44 meV and a rather small Stokes shift of 8 meV from the absorption line, which is typical for CdSe NPLs\cite{Ithurria2011nm,Tessier2012}. Representative spectra for the 3ML and 5ML samples are given in  Fig.~\ref{fig:SI_PL_Absorption_5ML_3ML_RT} and the corresponding parameters are listed in Table~\ref{tab:table1}.

\begin{figure*}[h!]
	\centering
	\includegraphics[width=16cm]{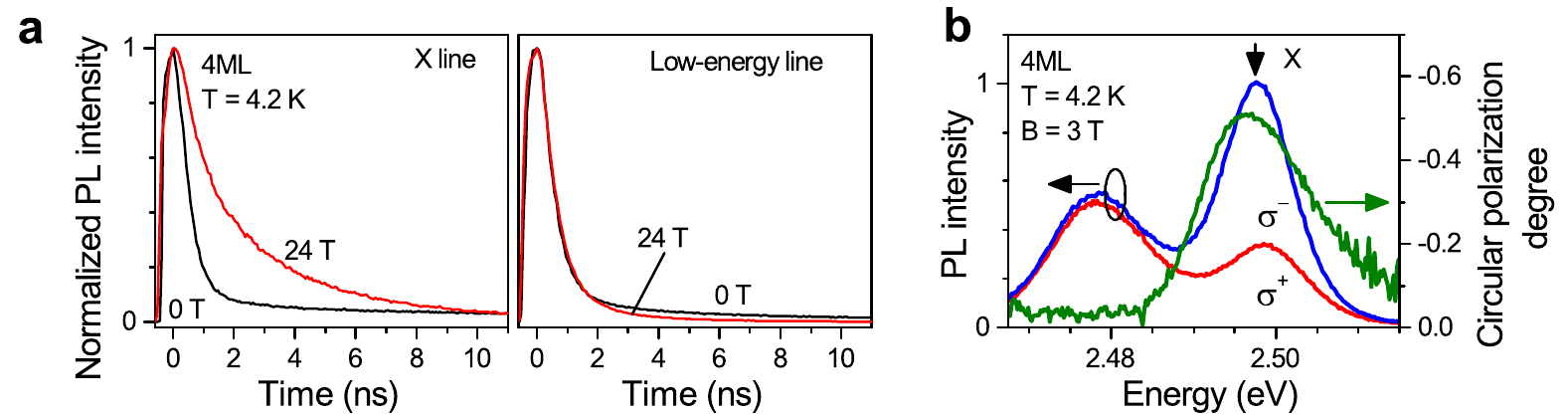}
	\caption{(a) PL decays of 4ML sample measured at $B=0$ (black) and $24$~T (red). Left panel: at the exciton line the PL decay is affected by magnetic field: the short component vanishes. Right panel: at the low-energy line the decay is not affected by magnetic field. (b) Intensity of $\sigma^-$ (blue) and $\sigma^+$ (red) circularly polarized PL and DCP (green) of 4ML sample in $B=3$~T and $T = 4.2$~K.}
	\label{fig:TR-PL(B)}
\end{figure*}

Figure~\ref{fig:PL}b shows PL spectra of all studied samples at $T=4.2$~K. The spectra consist of two lines, with the high-energy one (X) attributed to the exciton emission. The shift between these lines varies from 18 to 30 meV (Table~\ref{tab:table1}). The PL dynamics of the exciton line measured at $T=4.2$~K with an avalanche photodiode (APD) (Experimental section) is shown in Fig.~\ref{fig:PL}c. The decays exhibit the bi-exponential behavior typical for excitons in colloidal NCs, where the short decay is associated with bright exciton recombination and exciton relaxation from the bright to the dark state, while the long decay is associated with the dark exciton emission.
Monoexponential fits of the long-term tails are shown by the black lines, the corresponding decay times $\tau_{\rm L}$ range from 46 to 82~ns (Table~\ref{tab:table_model1}). In order to resolve the fast initial dynamics in the time range $20-30$~ps, a streak-camera detection was used (Experimental section). These results for the 4ML and 5ML samples are shown in Fig.~\ref{fig:PL}d together with exponential fits (the resulting times $\tau_{\rm short}$ are given in Table~\ref{tab:table_model1}), while the streak-camera images are shown in Fig.~\ref{fig:SI_Streak-camera}.  Thicker NPLs have a longer $\tau_{\rm short}$, the same trend was reported for spherical QDs\cite{Hannah2011}.

The origin of the low-energy line in the NPL emission spectra at low temperatures (Fig.~\ref{fig:PL}b) is still under debate.\cite{Tessier2013,Achtstein2016,Erdem2016phchl} Among the considered options are LO-phonon assisted exciton recombination,\cite{Tessier2013} emission of charged excitons (trions)\cite{Tessier2013} and that this line arises from recombination of a ground exciton state.\cite{Achtstein2016}
Several experimental features of the studied CdSe NPLs are in favor of the charged exciton origin of this low-energy line:
(i) The low-temperature absorption peak is close to the high-energy emission line proving its assignment to the exciton ground state (Fig.~\ref{fig:SI_PL_Absorption_5ML_5K}).
(ii) The energy separation between PL lines changes with NPL thickness and becomes larger than the 25~meV reported for the LO phonon energies in CdSe NPLs.\cite{Cherevkov2013,Dzhagan2016} (Table~\ref{tab:table1}) (iii) The recombination dynamics and its modification in external magnetic field are very different for the two lines. As one can see in the left panel of Fig.~\ref{fig:TR-PL(B)}a, the exciton decay of the high-energy line strongly changes in high magnetic fields of 24~T. Namely, its fast decay component becomes considerably longer and the long decay component shortens, which is a result of magnetic field mixing of the bright and dark exciton states.\cite{Efros1996,Liu2013} No effect of the magnetic field is found for the dynamics of the low-energy line (right panel of Fig.~\ref{fig:TR-PL(B)}a), which is typical for  charged excitons with a bright ground state.\cite{Liu2013}
(iv) Also the magnetic-field-induced degree of circular polarization (DCP) of the PL is very different for the emission of the high-energy and low-energy lines (Fig.~\ref{fig:TR-PL(B)}b) evidencing their different origins. The DCP is defined as $P_c = (I^+ - I^-)/(I^+ + I^-)$, where $I^+$ and $I^-$ are the intensities of the $\sigma^+$ and $\sigma^-$ circularly polarized emission, respectively. It is controlled by the Zeeman splitting of the exciton complexes and by their spin relaxation dynamics.\cite{Liu2013} The detailed analysis of the DCP goes beyond the scope of this paper and will be published elsewhere. In this paper we focus on the properties of the high-energy exciton emission line to investigate the fine structure of the neutral exciton in CdSe NPLs.

\subsection{Fluorescence line narrowing}

\begin{figure*}[h!]
	\centering
	\includegraphics[width=16cm]{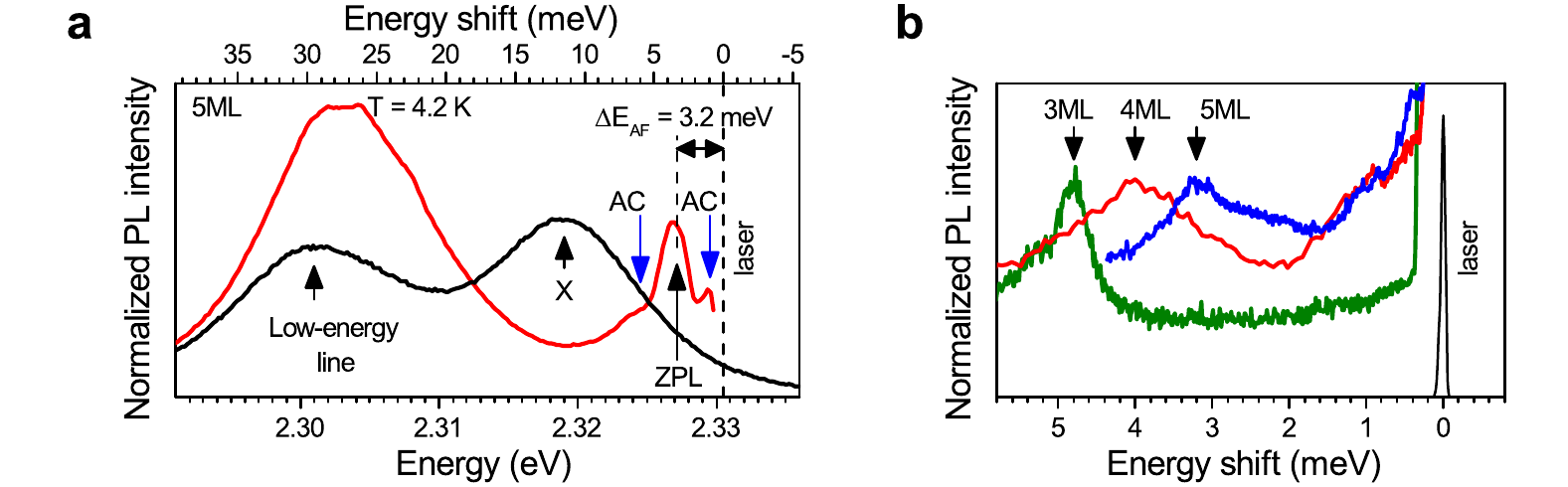}
	\caption{(a) Fluorescence line narrowing spectrum of 5ML sample for laser excitation at $2.3305$~eV (red) and PL spectrum under non-resonant excitation (black). Acoustic phonon replicas (AC) in the FLN spectrum associated with the bright and dark exciton states are marked by the blue arrows. (b) Fluorescence line narrowing spectra of 3ML, 4ML, and 5ML samples at laser excitation energies of $2.8076$, $2.5407$, and $2.3305$~eV, respectively.}
	\label{fig:FLN_5ML}
\end{figure*}

FLN is a commonly used technique to study the band edge exciton fine structure in colloidal NCs. It is technically demanding as it requires lasers, which photon energy can be tuned to the exciton resonances, and double or triple spectrometers with high suppression of the scattered laser light for measurements in the vicinity of the laser photon energy. FLN is used to resolve spectral lines in an inhomogeneously broadened ensemble by selective laser excitation.\cite{Nirmal1995,Furis2006,Wijnen2008,GranadosDelAguila2014,Biadala2017nn} Under resonant laser excitation within the inhomogeneously broadened exciton line, a subensemble of NPLs is selectively excited. This results in a strong narrowing of the emission lines in the PL spectrum, as the laser line is in resonance with the bright $\ket{\pm 1}$ exciton state of only a small fraction of NPLs. The injected excitons relax into the dark state $\ket{\pm 2}$, where the radiative recombination occurs. The Stokes shift between the laser photon energy and the dark exciton emission directly gives $\Delta E_{\rm AF}$ if possible contributions by dangling bond magnetic polarons or acoustic phonon polarons are absent\cite{Biadala2017nn} and internal relaxation between the exciton states can be neglected.

Figure~\ref{fig:FLN_5ML}a shows a PL spectrum of the 5ML sample under nonresonant excitation (black) and an FLN spectrum under resonant excitation at 2.3305~eV (red). In FLN experiment the broad exciton emission line marked as X vanishes. Instead, the FLN spectrum consisting of several lines appears. We attribute the line with the highest PL intensity to the zero-phonon line (ZPL). Its Stokes shift from the laser photon energy gives $\Delta E_{\rm AF}=3.2$~meV.  We assign two side peaks in vicinity of the ZPL line to acoustic phonon replicas of the dark and bright excitons. The results for all samples are shown in Fig.~\ref{fig:FLN_5ML}b. The bright-dark exciton splitting,  $\Delta E_{\rm AF}$, varies from 3.2~meV in the 5ML NPLs to 4.8~meV in the 3ML ones being about inversely proportional to the NPL thickness $L$ (Table~\ref{tab:table3}).

\subsection{Temperature-dependent time-resolved PL}

\begin{figure*}[h!]
	\centering
	\includegraphics[width=16cm]{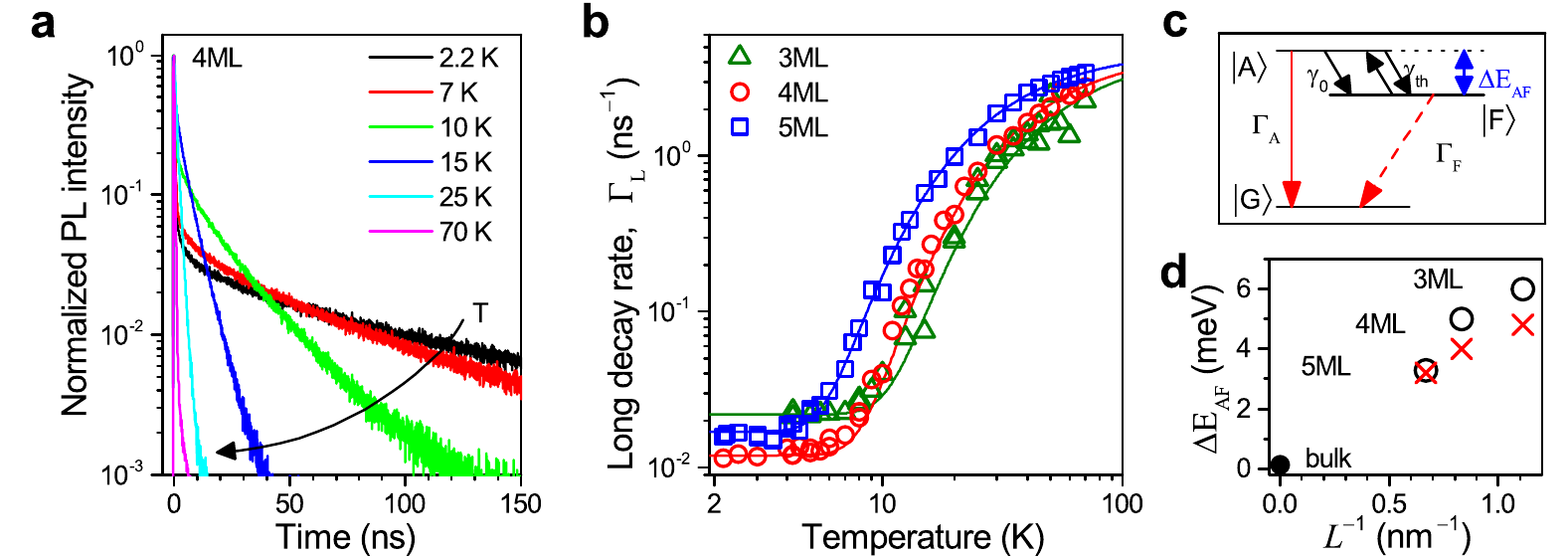}
	\caption{(a) PL decays at maxima of exciton line of 4ML sample measured at various temperatures. (b) Long component decay rate $\Gamma_{\rm L} = \tau_{\rm L}^{-1}$ as function of temperature. Lines are fits with equation~\eqref{eq:tauLongFull}, the resulting $\Gamma_{\rm A}$, $\Gamma_{\rm F}$ and $\Delta E_{\rm AF}$ are given in Table~\ref{tab:table_model1}. (c) Three-level model: $\ket{A}$ and $\ket{F}$ are bright and dark exciton states, and $\ket{G}$ is unexcited crystal state. (d) Bright-dark splitting $\Delta E_{\rm AF}$ \textit{vs} inverse sample thickness $L^{-1}$ measured by FLN (red crosses) and temperature-dependent time-resolved PL (black open circles). Closed circle is for bulk w-CdSe where $\Delta E_{\rm AF}^{\rm w}=0.13$~meV.\cite{Kiselev1975,Kochereshko1983}}
	\label{fig:TR_T}
\end{figure*}

The PL decays of the exciton lines, which are bi-exponential at liquid helium temperature, change with increasing temperature (Fig.~\ref{fig:TR_T}a): in 4ML $\tau_{\text{L}}$ drastically shortens from 82~ns at 2.2~K to 0.36~ns at 70~K, and the short decay component decreases in amplitude and vanishes for $T>30$~K. The recombination rates $\Gamma_{\rm L}=\tau_{\rm L}^{-1}$ deduced from monoexponential fits to the long-term tails for all samples are shown in Fig.~\ref{fig:TR_T}b as functions of temperature. Within the three-level model illustrated in Fig.~\ref{fig:TR_T}c (see Supplementary Section~S1 for more details), the recombination rate of the dark exciton, $\Gamma_{\rm F}$, is assumed to be temperature independent, and the acceleration of $\Gamma_{\rm L}$ with temperature is determined solely by the thermal population of the bright exciton state with the recombination rate $\Gamma_{\rm A} \gg \Gamma_{\rm F}$. The short component of the PL decay is determined by bright exciton recombination and exciton relaxation from the upper lying bright to the dark state with a rate $\gamma_0 (1+N_{\rm B})$, where $\gamma_0$ is the zero temperature relaxation rate, $N_{\rm B} = 1/ \left[ \exp {(\Delta E_{\rm AF} / kT)} -1 \right]$ is the Bose-Einstein phonon occupation (Fig.~\ref{fig:TR_T}c). This process requires a spin-flip of either the electron or the hole spin in the exciton and $\gamma_0$ is often referred to as a spin-flip rate. $\gamma_{th}=\gamma_0N_{\rm B}$ is the thermal-activation rate for the reversed process. Within the model, the rate equations for the populations of the bright and dark exciton states, $p_{\rm A}$ and $p_{\rm F}$, are:
\begin{eqnarray}
\frac{d p_{\rm A}}{dt} & = & -\left[\Gamma_{\rm A}+\gamma_0 (N_{\rm B}+1) \right]p_{\rm A} + \gamma_0 N_{\rm B} p_{\rm F} \nonumber , \\
\frac{d p_{\rm F}}{dt} & = & -\left[\Gamma_{\rm F}+\gamma_0 N_{\rm B} \right]p_{\rm F} + \gamma_0 (N_{\rm B}+1) p_{\rm A} . \label{eq:rate-equations}
\end{eqnarray}
Assuming $p_{\rm A}(t=0)=p_{\rm F}(t=0)=0.5$, the dependence of the decay rates on temperature is deduced from the solutions of rate equations \eqref{eq:rate-equations}:\cite{Biadala2009}
\begin{equation}
\begin{aligned}
\Gamma_{\rm short, L}(T) ={} & \frac{1}{2} \left[ \Gamma_{\rm A} +\Gamma_{\rm F}+\gamma_0 \coth\left( \frac{\Delta E_{\rm AF}}{2kT} \right) \pm \right. \\
& \left. \pm \sqrt{{\left( \Gamma_{\rm A} -\Gamma_{\rm F}+\gamma_0 \right)}^2+\gamma_0^2 \sinh^{-2}\left(\frac{\Delta E_{\rm AF}}{2kT} \right)} \right] .
\end{aligned}
\label{eq:tauLongFull}
\end{equation}
Here the sign ``$+$'' before the square root corresponds to $\Gamma_{\rm short}=\tau_{\rm short}^{-1}$ and the sign ``$-$'' to $\Gamma_{\rm L}$.

At low temperatures, such that $\Delta E_{\rm AF}\gg kT$, $\Gamma_{\rm L}=\Gamma_{\rm F}$ and $\Gamma_{\rm short}=\Gamma_{\rm A}+\gamma_0$. Substituting $\gamma_0$ by $\gamma_0=\Gamma_{\rm short}(T=2\text{ K})-\Gamma_{\rm A}$, we fit the $\Gamma_{\rm L}(T)$ dependences in Fig.~\ref{fig:TR_T}b with the equation~\eqref{eq:tauLongFull} and obtain the values of $\Delta E_{\rm AF}$ and $\Gamma_{\rm A}$. All evaluated parameters are given in Table~\ref{tab:table_model1}.
The bright exciton recombination rates $\Gamma_{\rm A} \sim 10$~ns$^{-1}$ are in good agreement with the reported $\Gamma_{\rm A} =3.6$ and $5.5$~ns$^{-1}$ for CdSe NPLs.\cite{Biadala2014nl} Note, that $\Gamma_{\rm A}$ in CdSe NPLs is about two orders of magnitude faster than in CdSe spherical NCs.\cite{Crooker2003,Labeau2003, DeMelloDonega2006,Biadala2009}

The zero-temperature relaxation rates are $\gamma_0=35.6 \text{ and } 24$~ns$^{-1}$ for the 4ML and 5ML NPLs, respectively. For the 3ML sample an estimate $\gamma_0=40$~ns$^{-1}$ was made assuming that both $\tau_{\rm short}^{-1}$ and $\gamma_0$ increase in thinner NPLs.\footnote[2]{Since no data for $\tau_{\rm short}$ in 3ML is available, $\gamma_0$ is estimated by extrapolating thickness dependence.}

The $\Delta E_{\rm AF}$ values obtained from the fit for all studied NPLs are plotted in Fig.~\ref{fig:TR_T}d together with the FLN results \textit{vs} the inverse NPL thickness $L^{-1}$. For reference we show also the value for bulk wurzite CdSe (w-CdSe) by a closed circle: $\Delta E_{\rm AF}^{\rm w}=0.13$~meV.\cite{Kiselev1975,Kochereshko1983}  For all samples, FLN gives slightly smaller values, but the trend is the same. These measurements confirm our previous result for CdSe NPLs with $\Delta E_{\rm AF}$ of a few meV.\cite{Biadala2014nl} Remarkably, the values from Table~\ref{tab:table_model1} are sufficient not only for characterizing the PL dynamics, but also for modeling the temperature evolution of the PL spectra without any additional parameters (see Subsection~\ref{Sec:Tdep}).

\begin{table}[h!]
	\small
	\caption{\ Fitting parameters from Figures~\ref{fig:PL}c, \ref{fig:PL}d and \ref{fig:TR_T}b}
	\label{tab:table_model1}
	\begin{tabular*}{0.48\textwidth}{@{\extracolsep{\fill}}llll}
		\hline
		Sample & 3ML & 4ML & 5ML \\
		\hline
		$\tau_{\rm L}$ ($T = 4.2$~K), ns & 46 & 82 & 58 \\
		$\tau_{\rm short}$ ($T = 2$~K), ps & --\footnote[3] & 22 & 29 \\
		$\Delta E_{\rm AF}$, meV & $6.0 \pm 0.5$ & $5.0 \pm 0.5$ & $3.3 \pm 0.5$ \\
		$\Gamma_{\rm A}$, ns$^{-1}$ & 10 & 10 & 10 \\
		$\Gamma_{\rm F}$, ns$^{-1}$ & 0.022 & 0.012 & 0.017 \\
		$\gamma_0$, ns$^{-1}$ & 40\footnote[3] & 35.6 & 24 \\
		\hline
	\end{tabular*}
\end{table}

\subsection{Spectrally-resolved PL decay} 
\begin{figure*}[h!]
	\centering
	\includegraphics[width=16cm]{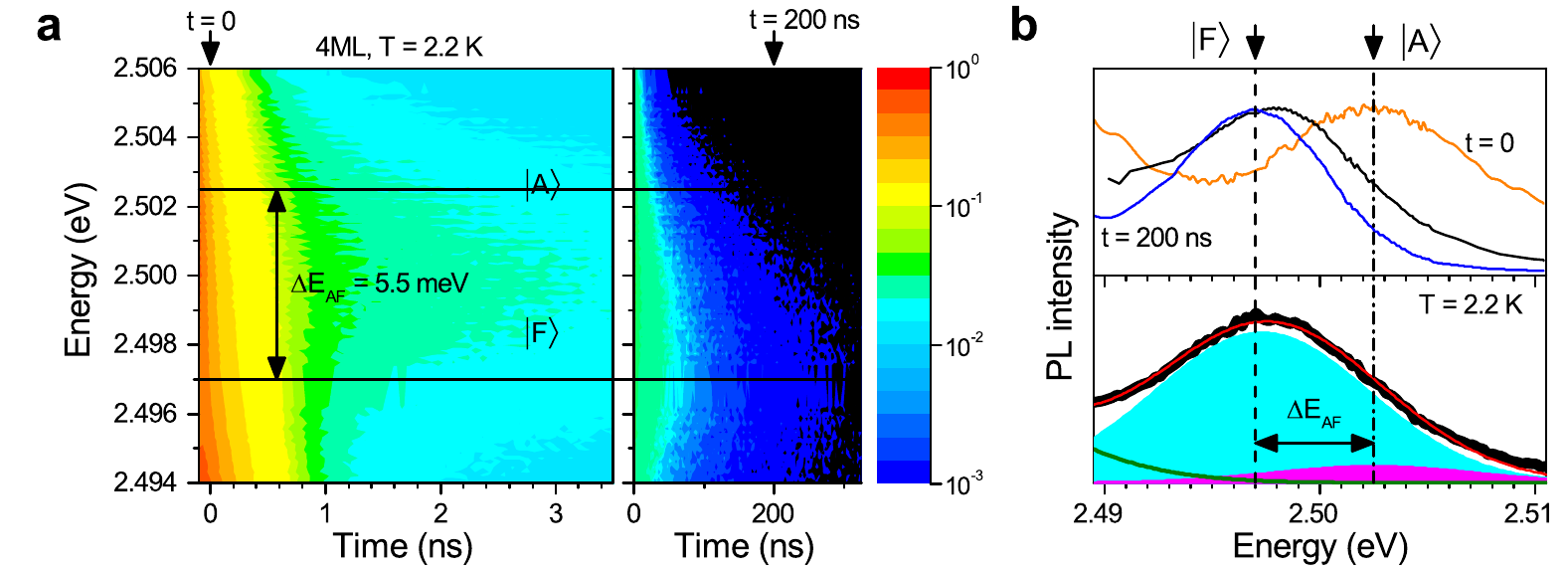}
	\caption{(a) Spectrally-resolved PL decays of 4ML sample at $T=2.2$~K for two temporal ranges. (b) Upper panel: PL spectra at different delays: $t=0$ (orange), $t=200$~ns (blue), and time-integrated over the laser repetition period (black).
		Lower panel: time-integrated PL spectrum at $T=2.2$~K measured with a CCD camera (black), fitted with three Gaussians corresponding to dark exciton (cyan), bright exciton (magenta) and low-energy peak (green). Red line gives the resulting curve. The fitting is the same as in Fig.~\ref{fig:Spe(T)}a.}
	\label{fig:PL_alternative_&_Quasi-streak}
\end{figure*}

To obtain more insight into the exciton emission of the NPLs, we performed a thorough analysis of the spectrally-resolved PL decay. Figure~\ref{fig:PL_alternative_&_Quasi-streak}a shows the time-resolved PL at different spectral energies (streak-camera-like data presentation) for the 4ML sample measured at $T=2.2$~K (Experimental section).

After absorption of a non-resonant laser pulse and exciton energy relaxation, the bright and dark excitons are populated about equally.\cite{Labeau2003} However, due to its much larger oscillator strength only the bright exciton contributes to the PL immediately after the laser excitation. An example of a time-resolved spectrum at $t=0$ is shown in Fig.~\ref{fig:PL_alternative_&_Quasi-streak}b (upper panel, orange). The emission line maximum is shifted to higher energy ($\sim 2.5025$~eV) compared to the time-integrated spectrum with the maximum at $2.498$~eV (black line). As the excitons relax towards thermal equilibrium, the bright state becomes depopulated, and the emission maximum shifts to lower energy. At a delay of $t=200$~ns the emission comes only from the dark exciton $\ket{F}$ state  with the emission line maximum  at $\sim 2.497$~eV (blue line). Therefore, the bright-dark splitting can be directly obtained from comparing the spectra at $t=0$ and $t \rightarrow \infty$. We obtain $5.5 \pm 0.5$~meV for the 4ML and $4.0 \pm 0.5$~meV for the 5ML NPLs (Fig.~\ref{fig:SI_5ML_Quasi-streak}). For comparison, the time-integrated spectrum measured with a CCD is shown in the lower panel of Fig.~\ref{fig:PL_alternative_&_Quasi-streak}b. The magenta and cyan Gaussian lines show the time-integrated contributions of the bright and dark excitons to the emission, respectively (see Subsection~\ref{Sec:Tdep}).

\subsection{Temperature dependence of PL spectra}\label{Sec:Tdep}

\begin{figure}[h!]
	\centering
	\includegraphics[width=8cm]{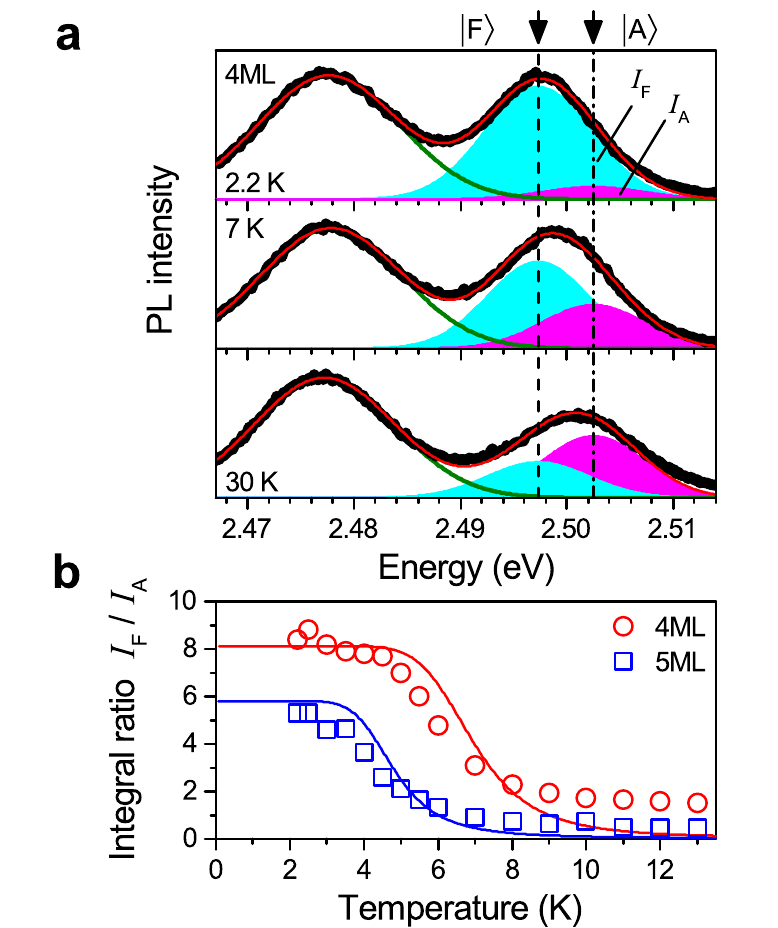}
	\caption{(a) PL spectra of 4ML sample at various temperatures. The spectra are fit with three Gaussian peaks (red line) with maxima corresponding to bright exciton (magenta filling), dark exciton (cyan filling), and low-energy peak (green line) positions.
		(b) Integral ratio $I_{\rm F}/I_{\rm A}$ derived from fitting the PL spectra at various temperatures (symbols). Lines are calculations with equation~\eqref{eq:integral_ratio} without fitting parameters.}
	\label{fig:Spe(T)}
\end{figure}

To explore in more detail the scattering rate between the dark and bright excitons and their splitting, $\gamma_0$ and $\Delta E_{\text{AF}}$, respectively, we analyzed the evolution of PL spectra with temperature for the 4ML (Fig.~\ref{fig:Spe(T)}a) and the 5ML (Fig.~\ref{fig:Spe(T)_5ML}) samples. At $T=2.2$~K (upper panel Fig.~\ref{fig:Spe(T)}a) the non-equilibrium exciton population relaxes into the lowest dark state and the maximum of the time-integrated exciton emission is at $\sim 2.4975$~eV. With increasing temperature (middle and lower panels) the population of the bright exciton state grows and the emission maximum shifts to higher energy. This behavior is in agreement with experiments on single NCs.\cite{Louyer2011}
To simulate the interplay between the exciton states, we fit the spectra with three Gaussian peaks centered at the energy positions corresponding to the bright exciton $E_{\rm A}$ (magenta filling), the dark exciton $E_{\rm F}$ (cyan filling), and the low-energy peak $E_{\rm LE}$ (green line). The full width at half maximum (FWHM) was kept fixed for all temperatures, and was $9.5$, $10$ and $13.3$ meV for the bright and dark exciton, and the low-energy peak, respectively (best fit). The fitting curves for PL spectra in Fig.~\ref{fig:Spe(T)}a are shown by the red lines. The best fit for the 4ML sample is achieved with $E_{\rm F}=2.4973$~eV and $E_{\rm A}=2.5025$~eV, which gives $\Delta E_{\rm AF}=E_{\rm A}-E_{\rm F}=5.2$~meV, in very good agreement with the results of temperature-dependent time-resolved measurements (Table~\ref{tab:table_model1}).

In addition to the energy splitting, the temperature dependence of the PL spectra brings insight into the thermal population of the bright and dark exciton states. Interestingly, within the three-level model described above, the integral PL intensity ratio of the dark to bright states, ${I_{\rm F}}/{I_{\rm A}}$, is directly linked to the bright-to-dark spin-flip rate, $\gamma_0$.
Integrating the set of equations \eqref{eq:rate-equations} and assuming $p_{\rm A}(0)=p_{\rm F}(0)=0.5$, we obtain
\begin{equation}
\frac{I_{\rm F}}{I_{\rm A}}=\frac{\Gamma_{\rm F}}{\Gamma_{\rm A}} \cdot \frac{\Gamma_{\rm A}+2 \gamma_0 (N_{\rm B}+1)}{\Gamma_{\rm F}+2 \gamma_0 N_{\rm B}}.
\label{eq:integral_ratio}
\end{equation}

Figure~\ref{fig:Spe(T)}b shows the experimental temperature dependence of the $I_{\rm F}/I_{\rm A}$ ratio and its calculation according to equation~\eqref{eq:integral_ratio} using the parameters from Table~\ref{tab:table_model1}. We stress, that good agreement is achieved without using any fitting parameters.

\section{Discussion}

\subsection{Bright-dark splitting}

In two-dimensional CdSe NPLs the exciton binding energy was estimated to amount 200--300~meV,\cite{Benchamekh2014} \textit{i.e.} in between the bulk CdSe value of 10 meV and the 500--1000 meV measured in 1--2 nm diameter CdSe QDs (Ref.~\onlinecite{Elward2013} and references therein). Therefore, $\Delta E_{\rm AF}$ values of the order of several meV, \textit{i.e.}  between the bulk (0.13 meV)\cite{Kiselev1975} and QD ($\sim 20$~meV)\cite{Nirmal1994} values are reasonable. They are about an order of magnitude larger than typical values in epitaxial II-VI QWs.\cite{Jeukens2002}
All four optical methods used in this paper provide consistent values of $\Delta E_{\rm AF}$ for the CdSe NPLs, which are collected in Table~\ref{tab:table3}.

\begin{table*}[h!]
	\small
	\caption{\ $\Delta E_{\rm AF}$ values in meV measured by different optical methods}
	\label{tab:table3}
	\begin{tabular*}{\textwidth}{@{\extracolsep{\fill}}llll}
		\hline
		Sample & 3ML & 4ML & 5ML\\
		\hline
		Fluorescence line narrowing & $4.8 \pm 0.1$  & $4.0 \pm 0.1$ & $3.2 \pm 0.1$ \\
		Temperature-dependent time-resolved PL & $6.0 \pm 0.5$ & $5.0 \pm 0.5$ & $3.3 \pm 0.5$  \\
		Spectrally-resolved PL decay & $-$ & $5.5 \pm 0.5$ & $4.0 \pm 0.5$ \\
		Temperature dependence of PL spectra & $-$  & $5.2\pm 0.5$ & $3.9 \pm 0.5$   \\
		\hline
	\end{tabular*}
\end{table*}

We would like to note here that the variety of optical methods presented in this paper can be further extended by application of magnetic fields. One example of such experiment is presented in Supplementary Section~S4. This method exploits the difference in the degree of circular polarization (DCP) of the bright and dark exciton emission in an external magnetic field due to the different Zeeman splittings, which is controlled by their $g$-factors. The DCP maximum indicates the position of the dark exciton (Fig.~\ref{fig:Spe(T)_1T}a), while the PL maximum shifts with increasing temperature from the dark to bright exciton position (Fig.~\ref{fig:Spe(T)_1T}b). The energy difference between the DCP and PL maxima of about 5~meV at $T>10$~K corresponds well with the $\Delta E_{\rm AF}$ values for the 4ML NPLs (Table~\ref{tab:table3}).

\subsection{Bright-dark splitting calculation within effective mass approximation, accounting for dielectric confinement effects}

The origin of the bright-dark splitting $\Delta E_{\rm AF}$ in NPLs is the electron-hole exchange interaction. Below we present calculations for $\Delta E_{\rm AF}$ obtained from consideration of the short-range exchange interaction. In the spherical approximation the exchange Hamiltonian can be written as:\cite{Efros1996}
\begin{eqnarray}
H_{\rm exch}= -\frac{2}{3}\varepsilon_{\rm exch}\nu\delta ({\bm r}_{\rm e}-{\bm r}_{\rm h})({\bm \sigma} \cdot{\bm J}),
\end{eqnarray}
where $\varepsilon_{\rm exch}$ is the exchange constant, ${\bm \sigma}=(\sigma_x,\sigma_y,\sigma_z)$ is the Pauli matrix, and $\bm J=(J_x,J_y,J_z)$ is the matrix of the hole total angular momentum $J=3/2$. Here we use the unit cell volume $\nu=\nu_{\rm c}=\textit{a}_{\rm c}^3$ (with $\textit{a}_{\rm c}$ being the lattice constant) for cubic material and $\nu=\nu_{\rm w}=\textit{a}_{\rm w}^2c_{\rm w}\surd 3/2$ (with $\textit{a}_{\rm w}$ and $c_{\rm w}$ being the lattice constants) for wurtzite semiconductors.

In NPLs strong confinement of the carriers occurs only in one direction so that exciton wavefunction can be written as:
\begin{eqnarray}
\Phi(\bm{r}_{\rm e},\bm{r}_{\rm h})=\Psi(\rho_{\rm e}-\rho_{\rm h})\psi(z_{\rm e})\psi(z_{\rm h}),
\end{eqnarray}
where $\Psi(\rho_{\rm e}-\rho_{\rm h})$ is the normalized wavefunction describing the exciton relative motion in the plane of a nanoplatelet, $\rho_{\rm e}$ and $\rho_{\rm h}$ are in-plane coordinates of electron and hole, respectively.  $\psi(z_{\rm e,h})=(2/L)^{1/2}\sin(\pi z_{\rm e,h}/L)$  is the wavefunction describing quantization of electron (hole) along the $z$ direction in an infinitely deep quantum well of thickness $L$.
The splitting between bright and dark excitons calculated using the wavefunction $\Phi(\bm{r}_{\rm e},\bm{r}_{\rm h})$ and the Hamiltonian $H_{\rm exch}$ gives:
\begin{eqnarray}
\Delta E_{\rm AF}=\Delta_{\rm exch}|\tilde \Psi(0)|^2/\tilde L,
\end{eqnarray}
where $\tilde L=L/\textit{a}_0$ is the dimensionless NPL thickness, $\tilde \Psi(0)=\Psi(0)\textit{a}_0$ is the dimensionless in-plane wavefunction evaluated at $\rho_{\rm e}=\rho_{\rm h}$, and $\Delta_{\rm exch}=\varepsilon_{\rm exch}\nu/\textit{a}_0^3$ is the renormalized exchange constant. Here we use $\textit{a}_0=1$~nm as the length unit.

The value of the renormalized exchange constant $\Delta_{\rm exch}$ is related to the bright-dark exciton spitting in bulk semiconductors as:\cite{Efros1996}
\begin{eqnarray}
\Delta_{\rm exch}^{\rm c}=\frac{3\pi}{8}\Delta E_{\rm AF}^{\rm c}\left( \frac{\textit{a}_{\rm ex}^{\rm c}}{\textit{a}_{0}}\right)^3, \\
\Delta_{\rm exch}^{\rm w}=\frac{\pi}{2}\Delta E_{\rm AF}^{\rm w}\left( \frac{\textit{a}_{\rm ex}^{\rm w}}{\textit{a}_{0}}\right)^3,
\end{eqnarray}
where $\textit{a}_{\rm ex}$ is the bulk exciton Bohr radius, "c" and "w" superscripts denote cubic and wurtzite material, respectively. This allows us to determine $\Delta_{\rm exch}^{\rm w}=35.9$~meV using the exciton splitting in w-CdSe $\Delta E_{\rm AF}^{\rm w}=0.13$~meV from Refs.~\onlinecite{Kiselev1975, Kochereshko1983} and the bulk exciton Bohr radius in w-CdSe $\textit{a}_{\rm ex}^{\rm w}=5.6$~nm\cite{Efros1996}.
As there is no experimental data for $\Delta E_{\rm AF}^{\rm c}$, we assume below that $\Delta_{\rm exch}^{\rm c}=\Delta_{\rm exch}^{\rm w}=35.9$~meV. The results of calculations for $\Delta E_{\rm AF}^{\rm c}$ with other possible choices of contributing parameters are given in Supplementary Section~S5.

To find $|\Psi(0)|^2$ we performed effective mass calculations for the exciton states following the approach from Refs.~\onlinecite{Gippius1998,Pawlis2011}. This approach includes the electron-hole Coulomb interaction and single particle potentials (Eqs.~(5) and (3) from Ref.~\onlinecite{Gippius1998}, respectively) modified by the difference in dielectric constants between the NPLs, $\epsilon_{\rm in}$, and the surrounding medium, $\epsilon_{\rm out}$. As electron and hole are localized inside a relatively small volume of the nanoplatelet, there arises a question: which dielectric constant $\epsilon_{\rm in}$ should be used for calculation of the Coulomb interaction between carriers inside the nanoplatelet? This issue has been raised previously \cite{Cardona,RodinaJETP2016} and concerns the number of resonances which give contribution to the dielectric response of the medium. We did the modeling for two values of $\epsilon_{\rm in}$ which equal to: (i) the high frequency dielectric constant of c-CdSe $\epsilon_{\infty}=6$,\cite{Benchamekh2014} which is relevant for the case when the quantum confinement energies of electron and hole are much larger than the energy of the optical phonon, and (ii) the background dielectric constant of CdSe $\epsilon_{\rm b}=8.4$, which takes into account the contribution from all crystal excitations except the exciton.\cite{Cardona} The value of the dielectric constant of the surrounding medium can vary in wide range, depending on the ligands at the NPL surface, the solvent, the substrate material on which the NPLs are deposited. Thus, we considered values of $\epsilon_{\rm out}$ ranging from 2, which is the case for randomly oriented ligands in solution\cite{Lechner1996,Benchamekh2014} (strong dielectric contrast), to $\epsilon_{\rm in}$ (dielectric contrast is absent). Here we present results of calculations with $\epsilon_{\rm in}=8.4$, $\epsilon_{\rm in}=6$ and $\epsilon_{\rm out}=2$.  For the results of calculations with other values of dielectric constants see Supplementary Section~S5.

One can see from Fig.~\ref{fig:Eaf_calculated_main_text} that calculations with $\Delta_{\rm exch}^{\rm c}=35.9$~meV and with dielectric constants $\epsilon_{\rm in}=6$, $\epsilon_{\rm out}=2$ or $\epsilon_{\rm in}=8.4$, $\epsilon_{\rm out}=2$ are in good agreement with the experimental data. We note that $\epsilon_{\rm in}=6$ and $\epsilon_{\rm out}=2$ also give a good agreement between the calculated and experimental absorption spectra of the CdSe NPLs \cite{Benchamekh2014}.
It is difficult to determine the exact values of dielectric constants $\epsilon_{\rm in}$, $\epsilon_{\rm out}$ and renormalized exchange constant $\Delta_{\rm exch}^{\rm c}$ in c-CdSe, as experimental data can be fitted using the wide range of these parameters (Supplementary Section~S5). However, all the parameterizations use reasonable set of fitting parameters $\epsilon_{\rm in}$, $\epsilon_{\rm out}$, $\Delta_{\rm exch}^{\rm c}$ and for all of them the calculation of $\Delta E_{\rm AF}$ based on the effective mass approximation with accounting for dielectric confinement effects agrees well with the experiment.

\begin{figure}[h!]
	\centering
	\includegraphics[width=8cm]{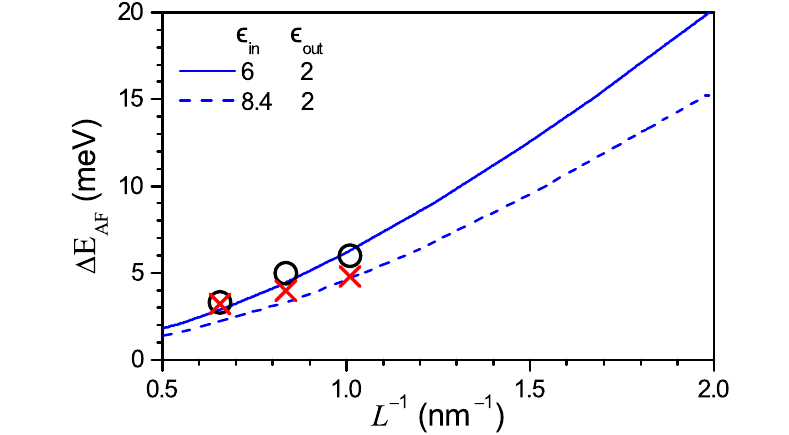}
	\caption{Dependence of $\Delta E_{\rm AF}$ on NPL thickness for the case of equal exchange constants in w-CdSe and c-CdSe. Lines show results of calculations with $\Delta_{\rm exch}^{\rm c}$=$\Delta_{\rm exch}^{\rm w}=35.9$~meV and $\epsilon_{\rm in}=6$, $\epsilon_{\rm out}=2$ (solid line) and $\epsilon_{\rm in}=8.4$, $\epsilon_{\rm out}=2$ (dashed line). Values of $\Delta E_{\rm AF}$ measured by FLN are shown by red crosses and values from temperature-dependent time-resolved PL are shown by black open circles.}
	\label{fig:Eaf_calculated_main_text}
\end{figure}

\subsection{Zero-temperature bright to dark relaxation rate}

We have shown by spectrally-resolved and time-resolved PL (Fig.~\ref{fig:PL_alternative_&_Quasi-streak}), that the bright excitons mostly contribute to the emission for $t< 500$~ps. Interestingly, even at a temperature as low as 2.2~K the PL signal from the bright exciton recombination still represents up to $10\%$ of the overall signal, as can be seen from the temperature dependence of PL spectra (Fig.~\ref{fig:Spe(T)}). This is due to the fact that, in contrast to spherical QDs where $\gamma_0 \gg \Gamma_{\rm A}$, in NPLs $\gamma_0\simeq25-40$~ns$^{-1}$ is only three times larger than $\Gamma_{\rm A}$ (Table~\ref{tab:table_model1}).
In small size QDs $\gamma_0$ of the same order of magnitude ($\sim 10$~ns$^{-1}$) were reported.\cite{Labeau2003}. This points to a considerably enhanced oscillator strength of the bright exciton in NPLs compared to QDs. Indeed, according to the present paper, in NPLs the bright exciton recombination rate $\Gamma_{\rm A}=10$~ns$^{-1}$, which is consistent with our previous measurement ($\Gamma_{\rm A} =3.6 $ and $5.5$~ns$^{-1}$, Ref.~\onlinecite{Biadala2014nl}), and is comparable with the one in epitaxial II-VI and III-V quantum wells under
nonresonant excitation.\cite{Feldmann1987,Polhmann1992}
In colloidal QDs $\Gamma_{\rm A}$ is about two orders of magnitude smaller: $0.082$\cite{Labeau2003}, $0.125$\cite{Crooker2003}, $0.025$\cite{DeMelloDonega2006}, and $0.16$\cite{Biadala2009}~ns$^{-1}$.
On the other hand, in epitaxially grown CdS quantum discs $\gamma_0=10$~ns$^{-1}$  and $\Gamma_{\rm A}=6$~ns$^{-1}$ have been reported.\cite{Gindele1998}  In this case, even large $\Delta E_{\rm AF}=4$~meV reported in these structures would not lead to prominent dark exciton emission, since the bright exciton decay would be dominated by radiative recombination rather then relaxation to the dark exciton. This raises the question of the impact of $\gamma_0$ onto emission properties of different nanostructures.

\section{Conclusions}

In summary, we have measured the parameters characterizing the band-edge excitons in CdSe nanoplatelets. We have used four optical techniques to study the exciton fine structure in ensembles of the nanoplatelets, in particular the bright-dark exciton splitting.  All techniques give consistent values for the bright-dark splitting $\Delta E_{\rm AF}$ ranging between $3.2$ and $6.0$~meV for the platelets thickness decreasing from 5 to 3 monolayers. The splitting scales about inversely with the platelet thickness. Theoretical calculations of $\Delta E_{\rm AF}$ based on the effective mass approximation with accounting for dielectric confinement effects were performed.
Despite of uncertainty of parameters and limited applicability of effective mass approximation for small-sized nanostructures, we find a good agreement between experimental and calculated size dependence of the bright-dark exciton splitting. The recombination rates of the bright and dark excitons and the bright to dark relaxation rate have been measured by time-resolved techniques. The recombination time of the bright excitons in nanoplatelets of about 100~ps is considerably faster than in colloidal QDs. As a result, in contrast to QDs with $\gamma_0 \gg \Gamma_{\rm A}$, in CdSe nanoplatelets $\gamma_0 \ge \Gamma_{\rm A}$, providing a different regime for the population of the bright and dark exciton states. A variety of the optical methods for measuring the bright-dark exciton splitting examined in this paper for CdSe colloidal nanoplatelets can be readily used for the whole family of colloidal nanostructures, which composition and design is in tremendous progress nowadays.






\section*{Experimental section}
\subsection*{Sample preparation}
The CdSe NPLs were synthesized according to the protocol reported in Ref.~\onlinecite{Ithurria2008}. They have a zinc-blend crystalline structure, i.e. c-CdSe. Samples for optical experiments were prepared by drop-casting of a concentrated NPL solution onto a quartz plate.

\subsection*{Optical measurements}
The optical experiments at low temperatures were performed on a set of different NPL ensemble samples. The NPL samples were mounted in a titanium sample holder on top of a three-axis piezo-positioner and placed in the variable temperature insert ($2.2-70$~K) of a liquid helium bath cryostat. For the measurements in external magnetic fields up to 17 T we used a cryostat equipped with a superconducting solenoid. For higher fields up to 24~T a cryostat was inserted in a 50~mm bore Florida-Bitter electromagnet at the High Field Magnet Laboratory in Nijmegen.
All optical experiments in magnetic fields were performed in Faraday geometry (light excitation and detection parallel to the magnetic field direction).

For nonresonant excitation measurements, the NPLs were excited using a pulsed diode laser (photon energy 3.06~eV, wavelength 405~nm, pulse duration 50~ps, repetition rate between 0.8 and 5~MHz) with a weak average excitation power density $<0.02$~W/cm$^2$.
The PL detected in backscattering geometry was filtered from the scattered laser light with a 0.55-m spectrometer and detected either by a liquid-nitrogen-cooled charge-coupled-device (CCD) camera or by an avalanche Si-photodiode. 
For polarization-resolved measurements, PL was analyzed by a combination of a quarter-wave plate and a linear polarizer. 
For the absorption spectra measurements at $T=5$~K the sample was illuminated by an incandescent lamp with a broad spectrum.

\subsection*{Time-resolved measurements with avalanche photodiode (APD)}
To measure long-lasting PL decays, we used an avalanche Si-photodiode connected to a conventional time-correlated single-photon counting setup (the instrumental response function is $\sim 100$ ps).

\subsection*{Spectral dependence of PL decay}
PL was filtered by a 0.55-m spectrometer equipped with a 2400 $\text{grooves}/\text{mm}$ grating, slicing the spectra into bands that were $\lesssim 1$~nm wide, and sent to an avalanche photodiode (APD). To prove that the APD quantum yield was the same for each wavelength range we compared the time-integrated PL spectrum with the PL spectrum measured by a the CCD camera.
To obtain the streak-camera-like image (Fig.~\ref{fig:PL_alternative_&_Quasi-streak}a) the time-resolved PL measured at different wavelengths was plotted across the energy in a two-dimensional plot. To obtain time-resolved PL spectra the two-dimensional data were integrated: for the spectrum at $t=0$ from $-32$ to $32$~ps and for the spectrum at $t=200$~ns from $195$ to $205$~ns.

\subsection*{Time-resolved measurements with a streak-camera}
In order to measure the initial fast PL dynamics, the NPLs were excited by a frequency-doubled mode-locked Ti-Sapphire laser (photon energy 3.06~eV, wavelength 405~nm, pulse duration 2~ps, repetition rate 76~MHz). Time-resolved PL spectra were recorded by a streak-camera attached to a spectrometer, providing temporal and spectral resolution of $\lesssim 5$~ps and $\lesssim 1$~nm. In these experiments the samples were in  contact with superfluid helium providing a temperature of about 2~K.

\subsection*{Fluorescence line narrowing}
For resonant excitation of the 5ML sample (Figure~\ref{fig:FLN_5ML}a) a continuous-wave laser with photon energy 2.3305~eV (wavelength 532~nm) was used. The signal was passed through a notch filter to suppress the scattered laser light. The PL was dispersed by a triple-grating Raman spectrometer (subtractive mode). The resonant PL emission was dispersed by a 500 mm stage (1800 grooves/mm holographic grating) and detected by a liquid nitrogen cooled CCD.

For excitation of 3ML, 4ML and 5ML samples (Figure~\ref{fig:FLN_5ML}b), we used the
lines of Ar-ion ($514.5$~nm, $486.5$~nm, $488$~nm), He-Cd ($441.6$~nm), and Nd:YAG ($532$~nm) lasers. The laser power densities focused on the sample was not higher than $2\text{ W}/\text{cm}^2$. The scattered light was analyzed by
a Jobin-Yvon U1000 double monochromator equipped with a cooled GaAs photomultiplier and conventional photon counting electronics.

\section*{Acknowledgements}
The authors are thankful to  Al.L. Efros for fruitful discussions. The authors are thankful to M. Meuris from department biomaterials and polymer science at TU Dortmund university for the TEM images of 4ML and 5ML samples.
E.V.S., V.V.B., D.R.Y., A.V.R., and M.B. acknowledge support of the Deutsche Forschungsgemeinschaft in the frame of ICRC TRR 160.
E.V.S. and D.R.Y. acknowledge the Russian Science Foundation (Grant No. 14-42-00015). N.A.G. acknowledges support from the Russian Foundation for Basic Research (Grant No. RFBR16-29-03283).
We acknowledge the support from HFML-RU/FOM, a member of the European Magnetic Field Laboratory (EMFL).
B.D. and Y.J. acknowledge funding from the EU Marie Curie project 642656 ``Phonsi''.

\balance

\onecolumn

\clearpage

\section*{Supplementary information:  \\ Addressing the exciton fine structure in colloidal \\ nanocrystals: the case of CdSe nanoplatelets}

\textit{Elena V. Shornikova, Louis Biadala, Dmitri R. Yakovlev, Victor F. Sapega, Yuri G. Kusrayev, Anatolie A. Mitioglu, Mariana V. Ballottin, Peter C. M. Christianen, Vasilii V. Belykh, Mikhail V. Kochiev, Nikolai N. Sibeldin, Aleksandr A. Golovatenko, Anna V. Rodina, Nikolay A. Gippius, Michel Nasilowski, Alexis Kuntzmann, Ye Jiang, Benoit Dubertret, and Manfred Bayer}

\setcounter{equation}{0}
\setcounter{figure}{0}
\setcounter{table}{0}
\setcounter{page}{1}
\renewcommand{\theequation}{S\arabic{equation}}
\renewcommand{\thefigure}{S\arabic{figure}}
\renewcommand{\thetable}{S\arabic{table}}
\renewcommand{\thepage}{S\arabic{page}}

\subsection*{S1. Band-edge exciton fine structure} \label{sec:X_Fine_Str}

It is well established theoretically and experimentally that in nanometer-sized colloidal semiconductor crystals the lowest eightfold degenerate exciton energy level is split into five fine structure levels by the intrinsic crystal field (in hexagonal lattice structures), the crystal shape asymmetry, and the electron-hole exchange interaction.\cite{Efros1996, Rodina2016} These levels are separated from each other by so large splitting energies, that at temperatures of a few Kelvin  the photoluminescence (PL) arises from the two lowest exciton levels. In nearly spherical CdSe wurtzite QDs\cite{Nirmal1995, Labeau2003, Crooker2003, DeMelloDonega2006, Biadala2010, Brovelli2011}, as well as in zinc blende NPLs\cite{Biadala2014nl}, the ground exciton state has total spin projection on the quantization axis $J=\pm 2$ and is forbidden in the electric-dipole (ED) approximation.\footnote[4]{In colloidal NCs, the exciton ground state is usually dark, with projection either $\pm 2$ or $0^{\rm L}$, depending on the shape and/or crystal structure.} Therefore, it is usually referred to as a ``dark'' state, $\ket{F}$. The upper lying ``bright'' state, $\ket{A}$, has $J=\pm 1^L$, and is ED allowed. The energy separation between these two levels $\Delta E_{\rm AF}=E_{\rm A}-E_{\rm F}$ is usually of the order of several meV and is relatively large compared to epitaxially grown quantum wells and quantum dots. These levels are schematically shown together with the relevant recombination and relaxation processes in Figure~\ref{fig:TR_T}c.

Typically, the linewidth of ensemble PL spectra of colloidal nanocrystals is one-two orders of magnitude larger than the characteristic $\Delta E_{\rm AF}=1-20$~meV. There are two optical methods that are commonly used to measure $\Delta E_{\rm AF}$ in different NCs.

\textbf{1. Temperature-dependent time-resolved PL} 

The exciton fine structure leads to an interplay between the upper lying bright $\ket{A}$ and the lower dark $\ket{F}$ states that is typical for colloidal nanostructures.
The recombination rates of these exciton states are $\Gamma_{\rm A}$ and $\Gamma_{\rm F}$. The PL intensity in this case can be written as $I(t)=\eta_{\rm A} \Gamma_{\rm A} p_{\rm A} + \eta_{\rm F} \Gamma_{\rm F} p_{\rm F}$, where $\eta_{\rm A,F}$ are the corresponding quantum efficiencies, and $p_{\rm A,F}$ are the occupation numbers of the corresponding levels. The relaxation rates between these levels are given by $\gamma_0$ and $\gamma_{th}$, where $\gamma_0$ is the zero-temperature relaxation rate, and $\gamma_{th}=\gamma_0 N_{\rm B}$ corresponds to the thermally-activated relaxation rate form the bright to the dark exciton state, where $N_{\rm B} = 1/ \left[ \exp ({\Delta E_{\rm AF} / kT}) -1 \right]$ is the Bose--Einstein phonon occupation. Assuming that $\gamma_0$, $\Gamma_{\rm A}$ and $\Gamma_{\rm F}$ are temperature independent parameters, the system dynamics can be described by the set of rate equations \eqref{eq:rate-equations}.
The solutions of this system are:
\begin{eqnarray}
p_{\rm A}&=&C_1 e^{-t  \Gamma_{\rm short}} + C_2 e^{-t  \Gamma_{\rm L}}, \nonumber \\
p_{\rm F}&=&C_3 e^{-t  \Gamma_{\rm short}} + C_4 e^{-t  \Gamma_{\rm L}},
\end{eqnarray}
with $\Gamma_{\rm short}=\tau_{\rm short}^{-1}$ and $\Gamma_{\rm L}=\tau_{\rm L}^{-1}$ being the rates for the short-lasting and the long-lasting decays, respectively:
\begin{equation}
\Gamma_{\rm short, L}(T)=\frac{1}{2} \left[ \Gamma_{\rm A} +\Gamma_{\rm F}+\gamma_0 \coth\left( \frac{\Delta E_{\rm AF}}{2kT} \right) \pm \sqrt{{\left( \Gamma_{\rm A} -\Gamma_{\rm F}+\gamma_0 \right)}^2+\gamma_0^2 \sinh^{-2}\left(\frac{\Delta E_{\rm AF}}{2kT} \right)} \right], \label{eq:tauLongFull2}
\end{equation}
Here the sign ``$+$'' in front of the square root corresponds to $\Gamma_{\rm short}$ and the sign ``$-$'' to $\Gamma_{\rm L}$.
For nonresonant excitation, after the laser pulse absorption, both $\ket{A}$ and $\ket{F}$ levels are assumed to be populated equally with $p_{\rm A}(t=0)=p_{\rm F}(t=0)=0.5$, which gives:
\begin{eqnarray}
p_{\rm A}&=&C_1 e^{-t  \Gamma_{\rm short}}+(0.5-C_1) e^{-t  \Gamma_{\rm L}}, \nonumber \\
p_{\rm F}&=&C_3 e^{-t  \Gamma_{\rm short}}+(0.5-C_3) e^{-t  \Gamma_{\rm L}}.
\end{eqnarray}
Here $C_1$ and $C_3$ are temperature dependent parameters:
\begin{eqnarray}
C_1&=&\frac{\gamma_0+\Gamma_{\rm A}-\Gamma_{\rm L}}{2(\Gamma_{\rm short}-\Gamma_{\rm L})}, \nonumber \\
C_3&=&\frac{-\gamma_0+\Gamma_{\rm F}-\Gamma_{\rm L}}{2(\Gamma_{\rm short}-\Gamma_{\rm L})}.
\end{eqnarray}

The PL intensity is then described by:
\begin{equation}
I(t)=\left[\eta_{\rm A} \Gamma_{\rm A} C_1 +\eta_{\rm F} \Gamma_{\rm F} C_3\right] e^{-t  \Gamma_{\rm short}}+ \left[\eta_{\rm A} \Gamma_{\rm A} (0.5-C_1) +\eta_{\rm F} \Gamma_{\rm F} (0.5-C_3)\right] e^{-t  \Gamma_{\rm L}}.
\end{equation}
This dependence represents a bi-exponential PL decay, as typically observed in colloidal NCs at cryogenic temperatures. Indeed, after nonresonant photoexcitation and energy relaxation of excitons the bright and dark states at $t=0$ are populated about equally, but only the emission from the bright exciton is observed due to $\Gamma_{\rm A} \gg \Gamma_{\rm F}$. In the limit $kT =0$, the excitons relax to the $\ket{F}$ state with a rate $\gamma_0$. These two processes, namely, recombination of the bright exciton and relaxation to the dark state, result in a fast initial drop of the time-resolved PL with a rate $\Gamma_{\rm short}=\Gamma_{\rm A}+\gamma_0(1+2N_{\rm B}) \approx \gamma_0(1+2N_{\rm B})$. At longer delays, the $\ket{A}$ level is emptied, and the emission arises from the $\ket{F}$ state with a rate $\Gamma_{\rm L}=\Gamma_{\rm F}$.

At a temperature of a few Kelvin, when $\Delta E_{\rm AF} \gg kT$ the time-resolved PL is also bi-exponential with the decay rates $\Gamma_{\rm short}$ and $\Gamma_{\rm L}$ defined by equation~\eqref{eq:tauLongFull2}. When the temperature is increased, the short-lived (long-lived) component decelerates (accelerates). If $\gamma_0 \gg \Gamma_{\rm A}$, at elevated temperatures corresponding to $\Delta E_{\rm AF} \le kT$ 
the decay turns into becoming mono-exponential with $\Gamma_{\rm L}=(\Gamma_{\rm A}+\Gamma_{\rm F})/2$ (see Figure~\ref{fig:TR_T}a).

The temperature dependence of the $\Gamma_{\rm L}$ rate is therefore a powerful tool to measure the $\Delta E_{\rm AF}$ value.
At a single dot level, it has been shown that the energy splitting obtained by this method is in excellent agreement with the energy splitting directly measured from the PL spectra and also with  theoretical calculations.\cite{Biadala2009} The analysis of the temperature dependence of the time-resolved PL decay is routinely used to evaluate $\Delta E_{\rm AF}$ in NCs.\cite{Crooker2003, DeMelloDonega2006, Biadala2010, Brovelli2011, Biadala2014nl, Biadala2016} However, this method is indirect and might be affected by thermal activation of trap states,\cite{Crooker2003, DeMelloDonega2006} surface dangling bonds,\cite{Rodina2015, Biadala2017nn} as well as contributions from higher energy states.\cite{Biadala2016}

It is important to note, that typically in colloidal quantum dots $\gamma_0 \gg \Gamma_{\rm A}$ so that the equations (\ref{eq:tauLongFull}) can be simplified:\cite{Labeau2003}
\begin{eqnarray}
\Gamma_{\rm short}&=& \Gamma_{\rm A} + \gamma_0(1+2N_{\rm B}) \approx \gamma_0(1+2N_{\rm B}) \label{eq:tauShort} , \nonumber \\
\Gamma_{\rm L}(T)&=& \frac{\Gamma_{\rm A} +\Gamma_{\rm F}}{2} - \frac{\Gamma_{\rm A}   -\Gamma_{\rm F}}{2}\tanh\left(\frac{\Delta E_{\rm AF}}{2kT}\right) .
\end{eqnarray}
However, this simplification cannot be used in case of NPLs, where as we have shown in this paper $\Gamma_{\rm A}$ can be comparable with $\gamma_0$.

\textbf{2. Fluorescence line narrowing} 

By exciting resonantly a small fraction of the NCs, the broadening due to the size distribution is drastically reduced and linewidths down to 300~$\mu$eV can be measured\cite{GranadosDelAguila2014}. However, this method neglects any internal relaxation between the exciton states.\cite{DeMelloDonega2006} Moreover, it was shown recently that the Stokes shift in bare core CdSe QDs can be also contributed by formation of dangling bond magnetic polarons.\cite{Biadala2017nn} The FLN technique, therefore, may  overestimate $\Delta E_{\rm AF}$.


\subsection*{S2. Sample characterization}\label{sec:Sample_characterization}

\begin{figure*}[h!]
	\centering
	\includegraphics[width=16cm]{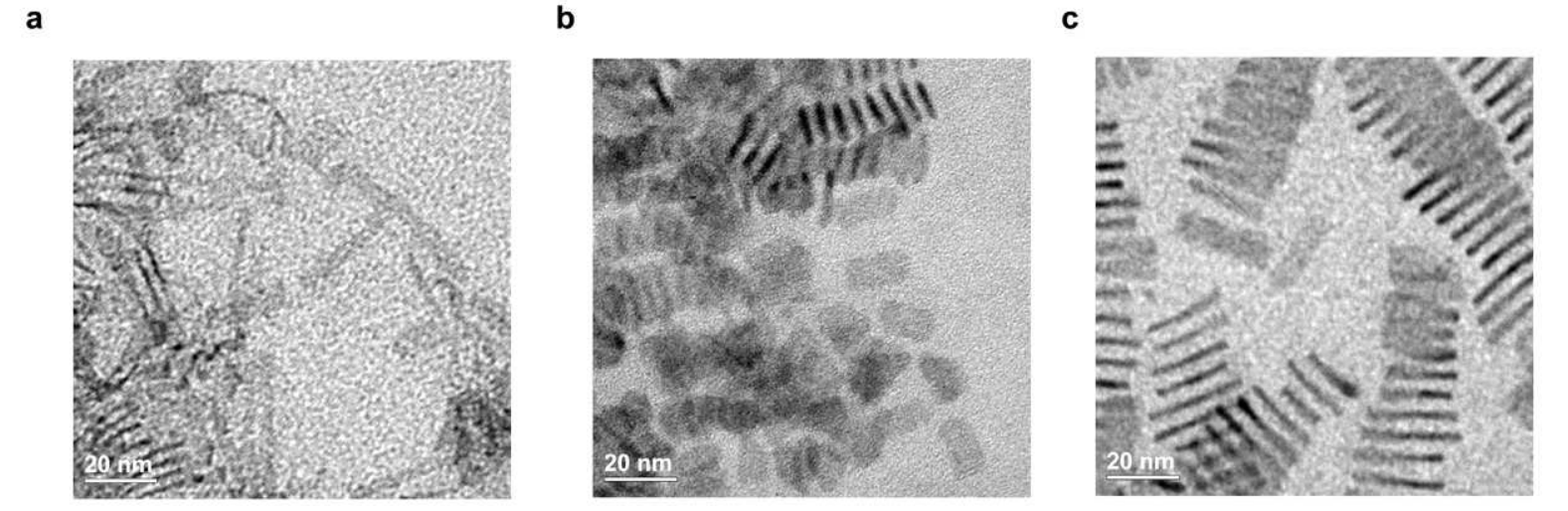}
	\caption{TEM images of (a) 3ML, (b) 4ML, (c) 5ML CdSe NPLs.}
	\label{fig:TEM}
\end{figure*}

\begin{figure*}[h!]
	\centering
	\includegraphics[width=15cm]{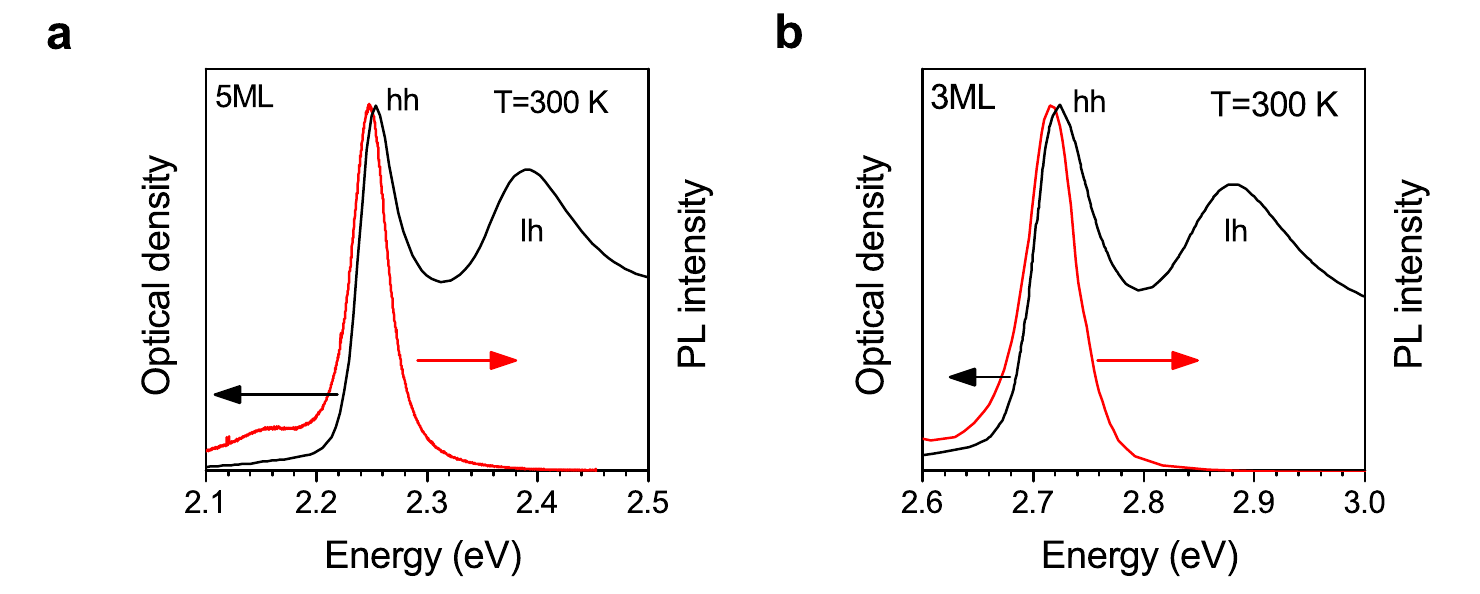}
	\caption{Emission (red) and absorption (black) spectra of (a) 5ML and (b) 3ML CdSe NPLs measured at $T=300$~K.}
	\label{fig:SI_PL_Absorption_5ML_3ML_RT}
\end{figure*}

\begin{figure*}[h!]
	\centering
	\includegraphics[width=8cm]{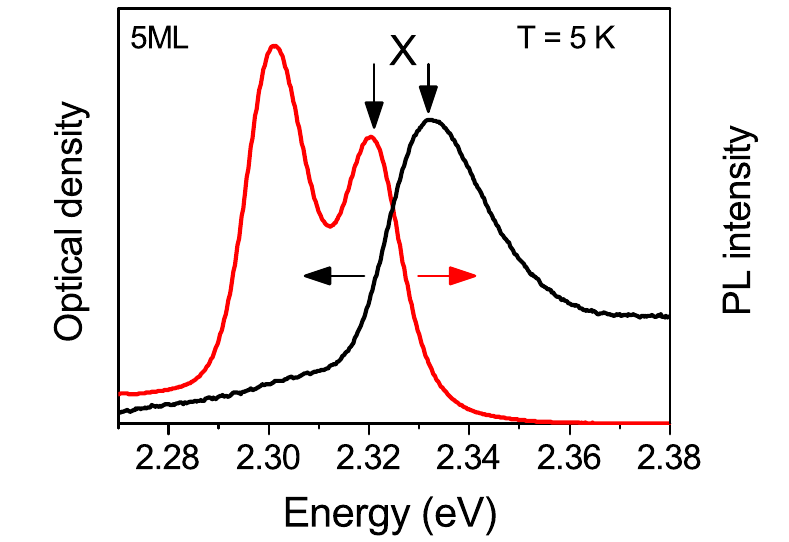}
	\caption{Emission (red) and absorption (black) spectra of 5ML CdSe NPLs at $T=5$~K. Exciton emission and absorption peaks are marked by arrows.}
	\label{fig:SI_PL_Absorption_5ML_5K}
\end{figure*}

\begin{figure*}[h!]
	\centering
	\includegraphics[width=15cm]{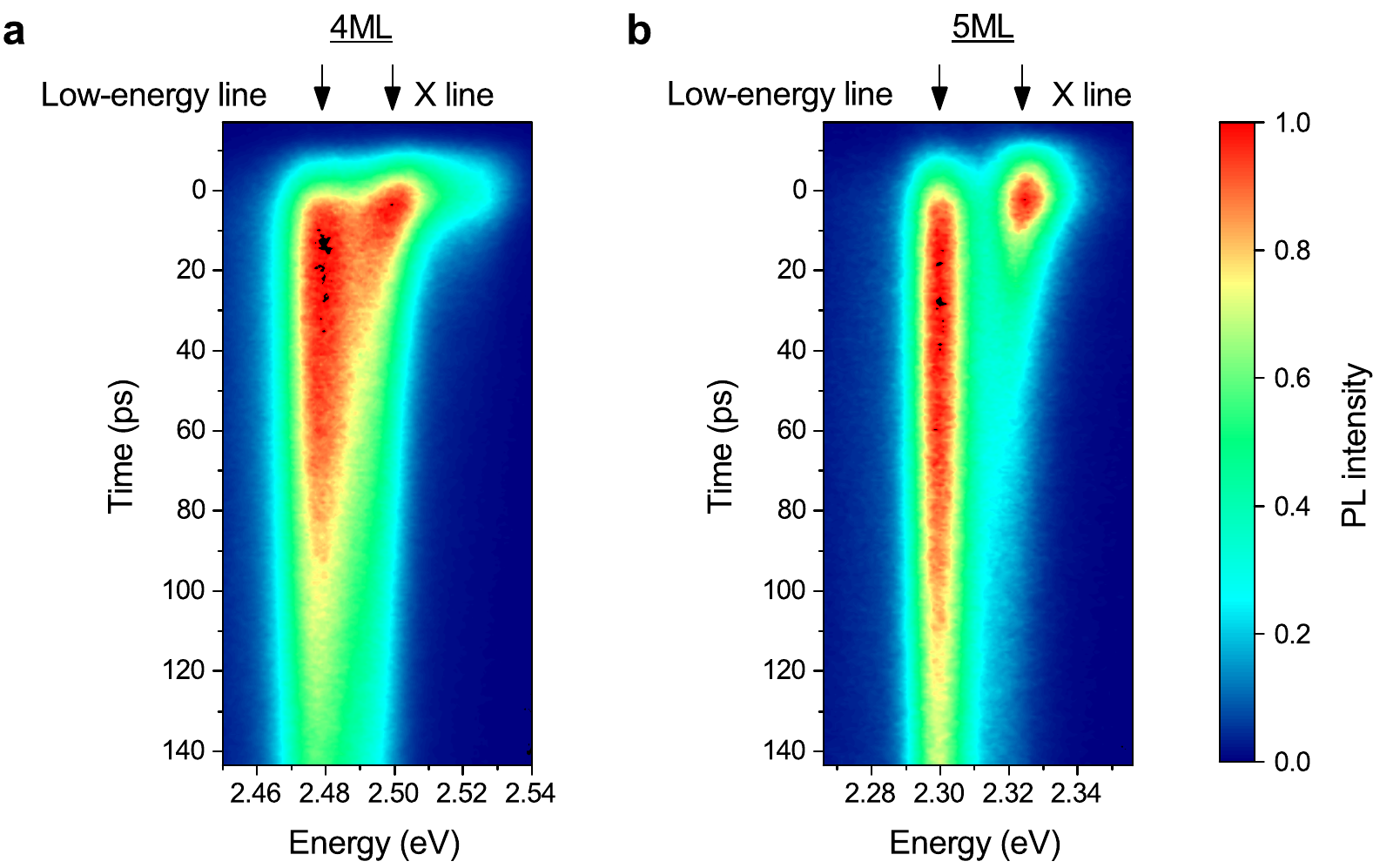}
	\caption{Evolution of exciton and low-energy line emission of 4ML and 5ML CdSe NPLs at $T=2$~K measured with a streak-camera.}
	\label{fig:SI_Streak-camera}
\end{figure*}

\clearpage
\subsection*{S3. Supplementary data for 5ML sample}\label{sec:Quasi-streak5ML}

\begin{figure*}[h!]
	\centering
	\includegraphics[width=15cm]{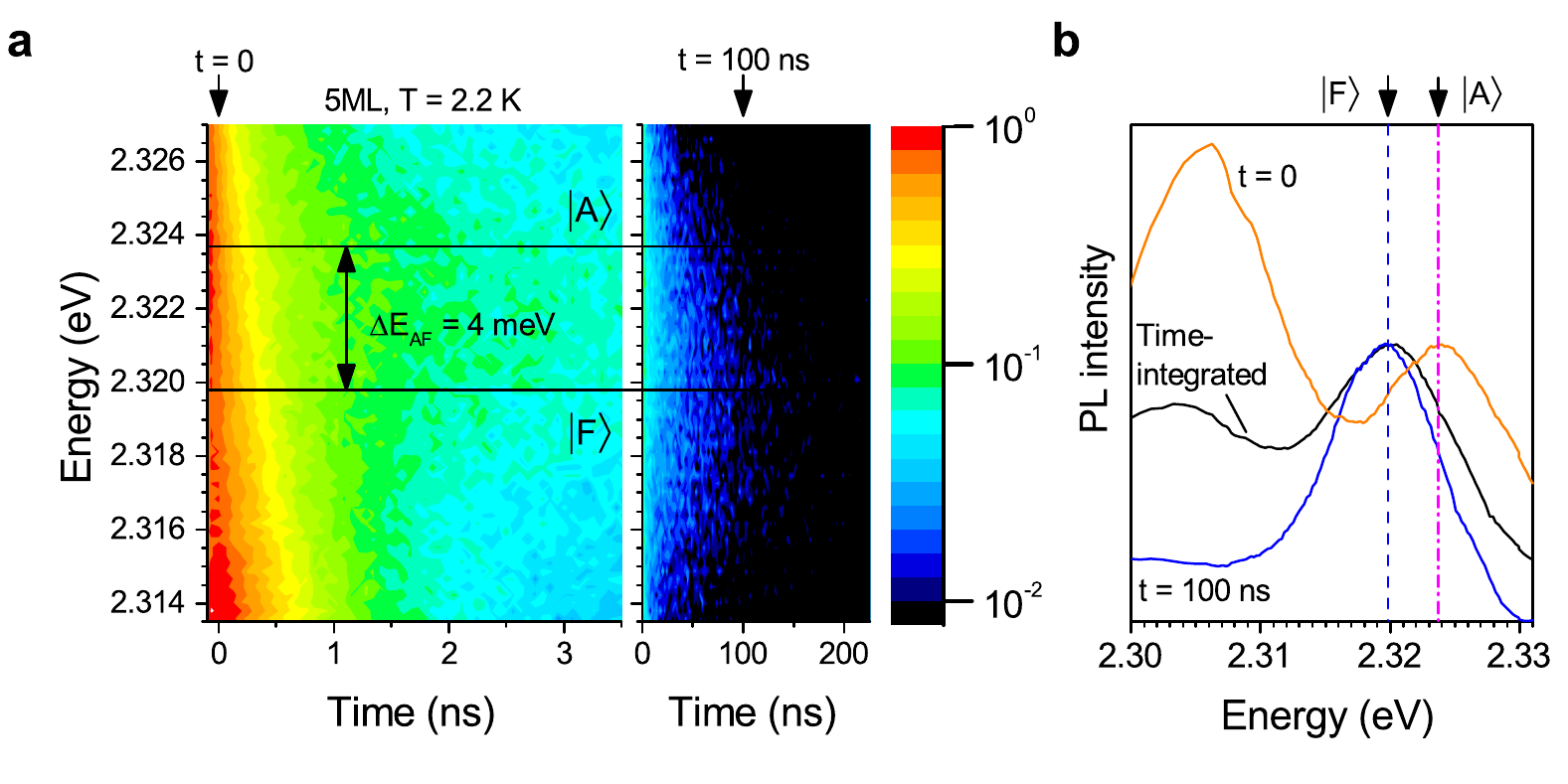}
	\caption{(a) Spectrally-resolved PL decays of 5ML sample at $T=2.2$~K shown for two temporal ranges. (b) PL spectra obtained by integration of the data in panel (a) over time at different delays: $t=0$ (orange, integration range $-32<t<32$~ps), $t=100$~ns (blue, integration range $95<t<105$~ns), and integrated over the whole period between subsequent laser pulses (black).}
	\label{fig:SI_5ML_Quasi-streak}
\end{figure*}

\begin{figure*}[h!]
	\centering
	\includegraphics[width=8cm]{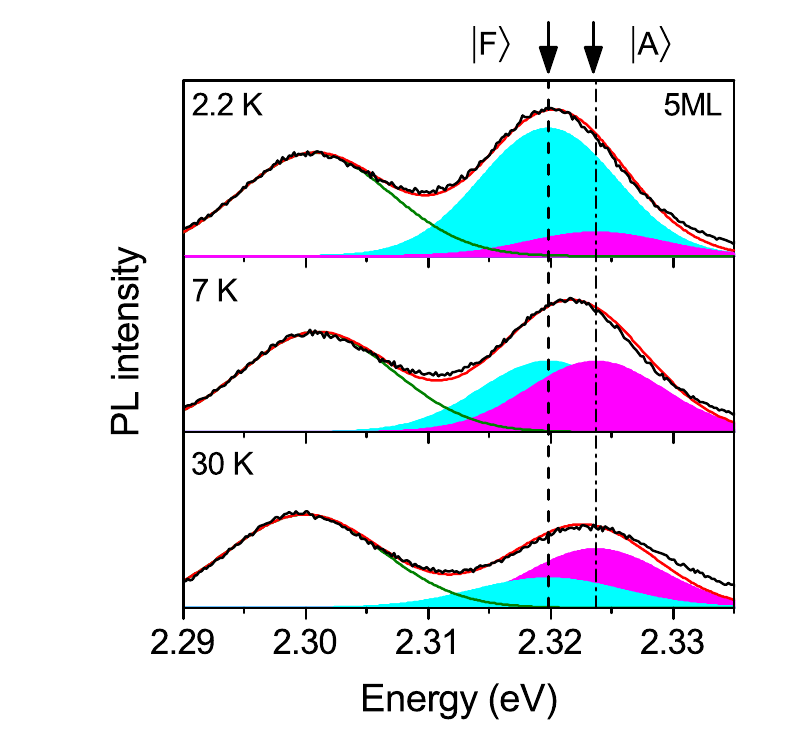}
	\caption{PL spectra of 5ML sample at various temperatures. The data are fit with three Gaussians with the peak maxima corresponding to the bright exciton (magenta), dark exciton (cyan), and low-energy peak (green) positions. The fit results for $I_{\rm F}/I_{\rm A}$ are presented in Figure~\ref{fig:Spe(T)}b.}
	\label{fig:Spe(T)_5ML}
\end{figure*}

\clearpage
\subsection*{S4. ``Method No. 5''. Polarization-resolved PL spectra in magnetic fields}
\label{sec:method5}

\begin{figure}[h!]
	\centering
	\includegraphics[width=8cm]{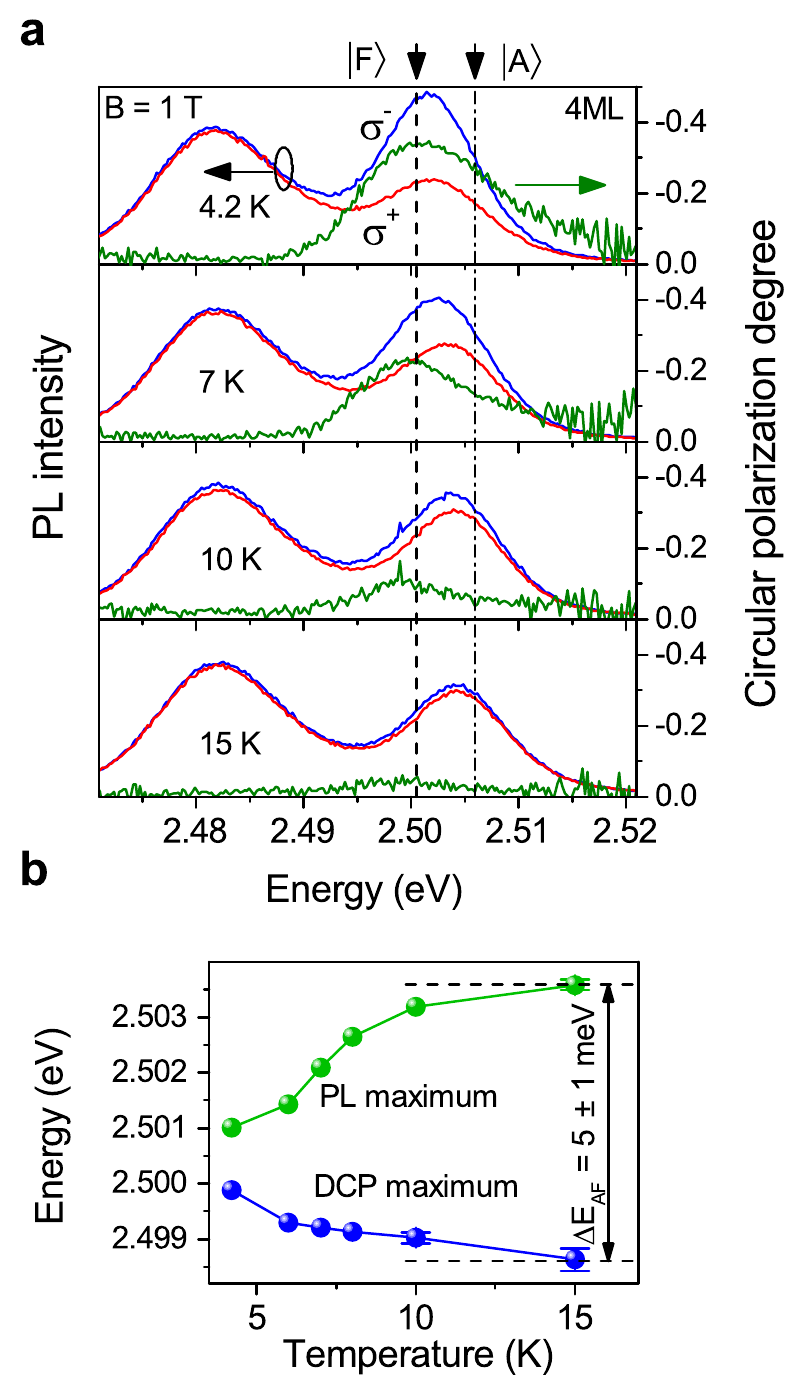}
	\caption{(a) PL spectra of 4ML sample at various temperatures measured at $B=1$~T. Left scale: intensity of $\sigma^-$ (blue) and $\sigma^+$ (red) circularly polarized PL components. Right scale: degree of circular polarization. The spectral position of DCP maximum indicates the dark exciton energy $E_{\rm F}$ and does not shift with temperature (black dashed line). 		(b) Spectral position of the DCP maximum (blue) and PL maximum (green) versus temperature.}
	\label{fig:Spe(T)_1T}
\end{figure}

The circularly polarized emission in an external magnetic field can be also used for identification of the bright and dark excitons in colloidal NPLs. This method exploits the difference in the Zeeman splittings of the bright and dark excitons, which is controlled by their $g$-factors, $g^A_X$ and $g^F_X$: $\Delta E_Z^{(A,F)}(B) = g^{(A,F)}_X \mu_B B \cos \theta$, where $\mu_B$ is the Bohr magneton and $\theta$ is the angle between the normal to the NPL plane and the magnetic field. Then the degree of circular polarization of the emission gained by the different thermal occupation of the exciton Zeeman sublevels is described by $P_c(B) = [\tau /(\tau + \tau_s)] \tanh [\Delta E_Z(B) /(2kT)]$. Here $\tau$ is exciton lifetime and $\tau_s$ is exciton spin relaxation time. The dark exciton state with angular momentum projection $\pm 2$ has $g$-factor $g^{F}_X = g_e-3g_h$.\cite{Efros1996,Efros2003} While the bright exciton state with $\pm 1$ has $g$-factor $g^{A}_X = -(g_e+3g_h)$ for the case when the exchange interaction is smaller than the splitting between the light-hole and heavy-hole states, which is valid for NPL. One can see, that the $g^{F}_X$ and $g^{A}_X$ can differ considerably. The difference depends on $g_e$ and $g_h$, which measurement for the studied NPLs goes beyond the scope of this paper.

A difference in $g$-factors has an immediate effect on the DCP by providing different values of $P_c(B)$ for the dark and bright excitons and different temperature dependences for them. This is confirmed by the experimental data in Fig.~\ref{fig:Spe(T)_1T}a, where the spectral dependence of the DCP is shown at $B=1$~T and various temperatures from 4.2 to 15~K. With increasing temperature the absolute value of DCP decreases, but its maximum remains located at the spectral position of the dark exciton, while the PL maximum shifts with increasing temperature from the dark to bright exciton position (Fig.~\ref{fig:Spe(T)_1T}b). The energy difference between the DCP and PL maxima of about 5~meV at $T>10$~K corresponds well with the $\Delta E_{\rm AF}$ values for the 4ML NPLs (Table~\ref{tab:table3}).

\clearpage
\subsection*{S5. Calculation of exciton parameters in c-CdSe NPL}
\label{sec:modeling}

In our calculations we consider only the contribution from the short-range exchange interaction to the bright-dark exciton splitting $\Delta E_{\rm AF}$ in c-CdSe NPLs. In spherical approximation it is described by:
\begin{eqnarray}
H_{\rm exch}= -\frac{2}{3}\varepsilon_{\rm exch}^{\rm c}\textit{a}_{\rm c}^3\delta ({\bm r}_{\rm e}-{\bm r}_{\rm h})({\bm \sigma} \cdot{\bm J}),
\end{eqnarray}
where $\varepsilon_{\rm exch}^{\rm c}$ is the exchange constant, $\textit{a}_{\rm c}=0.608$~nm is the lattice constant of c-CdSe \cite{Samarth}, ${\bm \sigma}=(\sigma_x,\sigma_y,\sigma_z)$ is the Pauli matrix, and $\bm J=(J_x,J_y,J_z)$ is the matrix of the hole total angular momentum $J=3/2$. We have found (see the main text) the resulting splitting as:
\begin{eqnarray}
\Delta E_{\rm AF}=\Delta_{\rm exch}|\tilde \Psi(0)|^2/\tilde L,
\end{eqnarray}
where $\tilde L=L/\textit{a}_0$ is the dimensionless NPL thickness, $\tilde \Psi(0)=\Psi(0)\textit{a}_0$ is the dimensionless in-plane wavefunction evaluated at $\rho_{\rm e}=\rho_{\rm h}$, and $\Delta_{\rm exch}=\varepsilon_{\rm exch}\nu/\textit{a}_0^3$ is the renormalized exchange constant. Here we use $\textit{a}_0=1$~nm as the length unit. 

The influence of dielectric contrast on the in-plane wavefunction of exciton $\Psi(0)$  is taken into account according to approach described in Ref.~[\onlinecite{Gippius1998}]. The full Hamiltonian of the system includes potential $U_{e,h}(\rho,z_e,z_h)$ which describes the Coulomb attraction between electron and hole, the attraction of the electron to the hole image, and of the hole to the electron image. Potential $U_{e,h}(\rho,z_e,z_h)$ depends on $\epsilon_{\rm out}$ and $\epsilon_{\rm in}$ as follows:  
\begin{eqnarray}
U_{e,h}(\rho, z_e,z_h)=-\frac{e^2}{\epsilon_{\rm in}}\left[\frac{1}{\sqrt{\rho^2+(z_e-z_h)^2}}+\frac{\epsilon_{\rm in}-\epsilon_{\rm out}}{\epsilon_{\rm in}+\epsilon_{\rm out}}\frac{1}{\sqrt{\rho^2+(z_e+z_h)^2}}\right ],
\end{eqnarray}
where $\rho=\rho_{\rm e}-\rho_{\rm h}$ is the exciton in-plane motion coordinate, $z_e$ and $z_h$ are coordinates of electron and hole along the quantization axis.  

Let us consider the results of the $\Delta E_{\rm AF}$ calculations, performed for different sets of dielectric constants of the nanoplatelet $\epsilon_{\rm in}$ and the surrounding media $\epsilon_{\rm out}$. We consider four different values of renormalized exchange constant $\Delta_{\rm exch}^{\rm c}$. 

\begin{figure*}[h!]
	\centering
	\includegraphics[width=16cm]{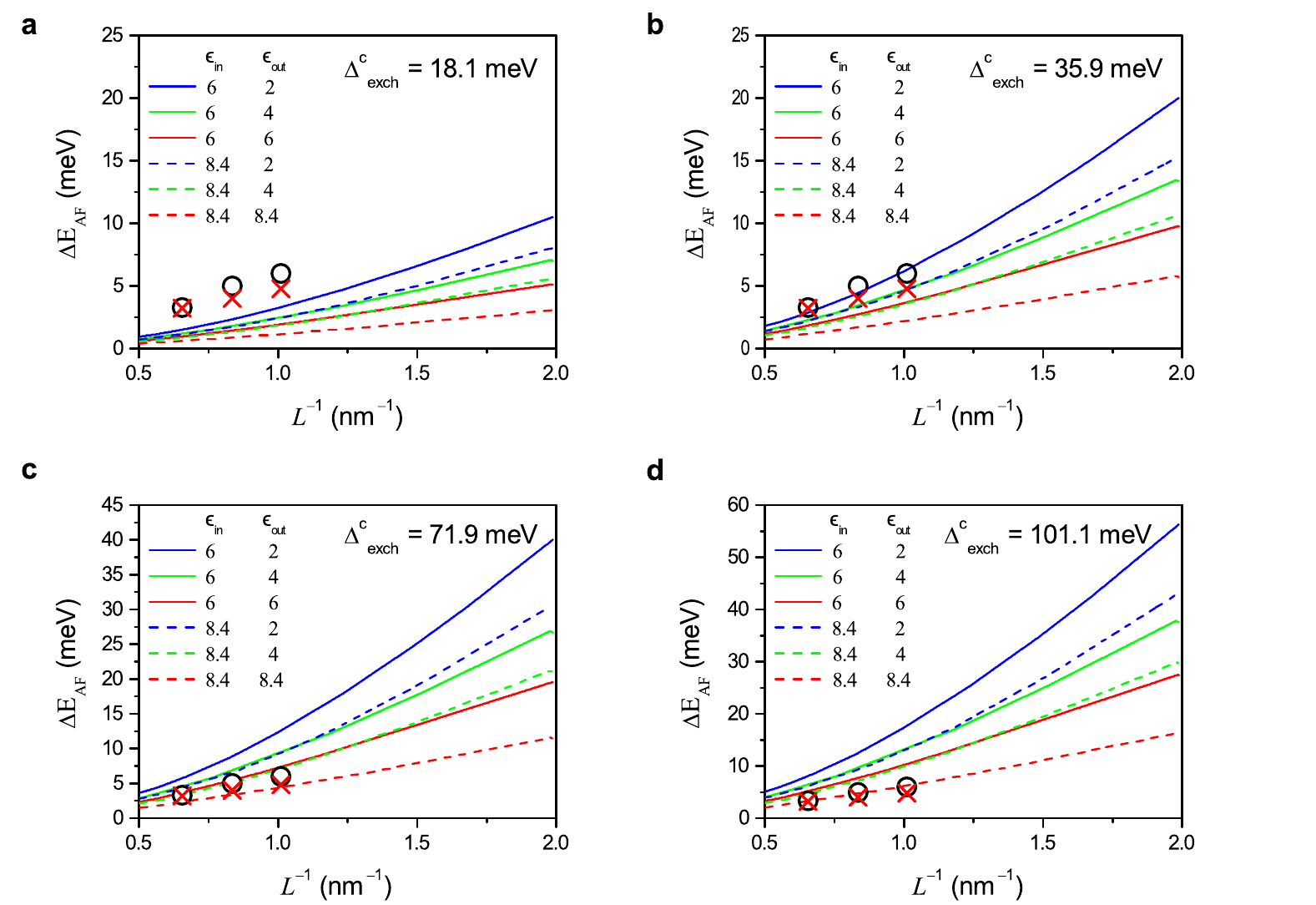}
	\caption{Dependence of $\Delta E_{\rm AF}$ on NPL thickness for: (a) $\Delta_{\rm exch}^{\rm c}=18.1$~meV, $\varepsilon_{\rm exch}^{\rm c}=84$~meV (b) $\Delta_{\rm exch}^{\rm c}=35.9$~meV, $\varepsilon_{\rm exch}^{\rm c}=160$~meV, (c) $\Delta_{\rm exch}^{\rm c}=71.9$~meV, $\varepsilon_{\rm exch}^{\rm c}=320$~meV, and (d) $\Delta_{\rm exch}^{\rm c}=101.1$~meV, $\varepsilon_{\rm exch}^{\rm c}=450$~meV. Lines are calculations. Values of $\Delta E_{\rm AF}$ measured by FLN are shown by red crosses and values from temperature-dependent time-resolved PL are shown by black open circles.}
	\label{fig:SI6_Eaf_calculations}
\end{figure*}

{\bf a.} The straightforward way to determine $\Delta _{\rm exch}^{\rm c}$ is based on the knowledge of the bright-dark splitting $\Delta E_{\rm AF}^{\rm c}$ in bulk c-CdSe (see Eq.7). However, there is no available experimental data for $\Delta E_{\rm AF}^{\rm c}$. The empirical expression for the bulk exchange splitting in zincblende semiconductors was obtained in Ref.~[\onlinecite{Fu1999}] from linear fit of splitting values in InP, GaAs and InAs. According to Eq.12 from Ref.~[\onlinecite{Fu1999}] we find:
\begin{eqnarray}
\Delta E_{\rm AF}^{\rm c}\left(\frac{\textit{a}^{\rm c}_{\rm ex}}{\textit{a}_{\rm 0}}\right)^3=15.4~{\rm meV}.
\end{eqnarray}
It corresponds to the renormalized exchange constant $\Delta_{\rm exch}^{\rm c}=18.1$~meV in c-CdSe, as well as in all other semiconductors with zincblende structure. One can see that this choice of $\Delta_{\rm exch}^{\rm c}$ gives calculated $\Delta E_{\rm AF}$ smaller than the experimental data at any $\epsilon_{\rm in}$ and $\epsilon_{\rm out}$ (Figure~\ref{fig:SI6_Eaf_calculations}a).

{\bf b.} The next approach is based on the assumption about equality of the renormalized exchange constants of c-CdSe and w-CdSe: $\Delta_{\rm exch}^{\rm c}=\Delta_{\rm exch}^{\rm w}=35.9$~meV. This approach gives good agreement with the experimental results if we use $\epsilon_{\rm in}$ varying from the high frequency dielectric constant of c-CdSe $\epsilon_{\infty}=6$ to the background dielectric constant of CdSe $\epsilon_{\rm b}=8.4$, and the outside dielectric constant $\epsilon_{\rm out}=2$. The results of calculations with the same $\Delta_{\rm exch}^{\rm c}$ and other sets of dielectric constants are presented in Fig.~\ref{fig:SI6_Eaf_calculations}b.

{\bf c.} Another approach is based on assumption about equality not of the renormalized exchange constants, but of the exchange constants $\varepsilon_{\rm exch}$ in c-CdSe and w-CdSe: $\varepsilon_{\rm exch}^{\rm c}=\varepsilon_{\rm exch}^{\rm w}=\Delta_{\rm exch}^{\rm w}{\textit{a}_0}^3/\nu_{\rm w}=320$~meV. Here $\nu_{\rm w}=\textit{a}_{\rm w}^2c_{\rm w}\surd{3}/2=0.112~\rm nm^3$  is the volume of the w-CdSe unit cell, \cite{Xu} where $\textit{a}_{\rm w}=0.43$~nm and $c_{\rm w}=0.70$~nm. Using the definition of c-CdSe unit cell from Refs.~[\onlinecite{Zarhri,Szemjonov}], we find $\nu_{\rm c}=\textit{a}_{\rm c}^3=0.224~\rm{nm}^3\approx 2 \nu_{\rm w}$, where $\textit{a}_{\rm c}=0.608~\rm{nm}$ according to Ref.~[\onlinecite{Samarth}], and $\Delta_{\rm exch}^{\rm c}=2\Delta_{\rm exch}^{\rm w}=71.9$~meV. The choices $\epsilon_{\rm in}=8.4$ and $\epsilon_{\rm out}=4$ fit the experimental data (Fig.~\ref{fig:SI6_Eaf_calculations}c). Note, that the value of $\Delta_{\rm exch}^{\rm c}=35.9$~meV in the case b corresonds to the $\varepsilon_{\rm exch}^{\rm c} = \Delta_{\rm exch}^{\rm c} a_0^3/\nu_{\rm c} = 160 meV$. 

{\bf d.} The last approach is also based on assumption about equality of the exchange constants $\varepsilon_{\rm exch}^{\rm c}=\varepsilon_{\rm exch}^{\rm w}$ with the use of $\varepsilon_{\rm exch}^{\rm w}=450$~meV from Ref.~[\onlinecite{Efros1996}].  It gives us $\Delta_{\rm exch}^{\rm c}=\varepsilon_{\rm exch}^{\rm c}\nu_{\rm c}/{\textit{a}_0}^3=101.1$~meV. The calculated $\Delta E_{\rm AF}$ is larger than the experimental data for any choice of $\epsilon_{\rm in}$ and $\epsilon_{\rm out}$, except of the not very realistic case without a dielectric contrast: $\epsilon_{\rm in}=\epsilon_{\rm out}=8.4$ (Figure~\ref{fig:SI6_Eaf_calculations}d).


While we can exclude the cases without dielectric confinement, when $\epsilon_{\rm in}=\epsilon_{\rm out}$, and the cases with $\Delta_{\rm exch}^{\rm c}<35.9$~meV, there are still a wide range of suitable parameterizations between those used in Figs.~\ref{fig:SI6_Eaf_calculations}~b,c. Independent determination of the renormalized exchange constant $\Delta_{\rm exch}^{\rm c}$, or dielectric constants $\epsilon_{\rm in}$, $\epsilon_{\rm out}$ would allow one to narrow down the number of parameterizations. However, all these parameterizations use reasonable values of $\epsilon_{\rm in}$, $\epsilon_{\rm out}$, $\Delta_{\rm exch}^{\rm c}$ and allow us to describe dependence of bright-dark exciton splitting in c-CdSe NPLs as a result of short-range exchange interaction between electron and hole within the effective mass approximation approach.

\end{document}